\newcommand{\pa}{\partial}
\newcommand{\be}{\begin{equation}}
\newcommand{\ee}{\end{equation}}
\newcommand{\bea}{\begin{eqnarray}}
\newcommand{\eea}{\end{eqnarray}}
\def \ci{\cite}
\def \YY {{\rm Y}}

\newcommand{\nn}{\nonumber}
\newcommand{\p}[1]{(\ref{#1})}
\newcommand{\bt}[1]{{\bar t}}

\newcommand{\ellK}{{\rm K}}

\newcommand{\ellE}{{\rm E}}

\def \cJ {{\cal J}}

\def \sm {{sigma-model}}

\newcommand{\sn}{\mathop{\mathrm{sn}}\nolimits}
\newcommand{\cn}{\mathop{\mathrm{cn}}\nolimits}
\newcommand{\dn}{\mathop{\mathrm{dn}}\nolimits}

\documentclass[12pt]{article}
\input epsf
\usepackage{epsfig,color,latexsym}
\def \sql {{\sqrt{\l}}\ }
\def \del{\partial}
\def \a {\alpha}
\def \aa {{\a'}}
\def\g{\gamma}
\def\s{\sigma}

\def\zi{\zeta_1}
\def\zii{\zeta_2}
\def\ov{\over}
\def\la{\label}

\def\J{{\cal J}}

\def \om {\omega}

\def\E{{\cal E}}
\def\w{\omega}
\def\b{\beta}
\def\l{\lambda}

\def \adss{$AdS_5 \times S^5$\ } 

\def \r { \rho}
\def \sql {\sqrt{\lambda} }
\def \t {\theta}
\def \p {\phi}
\def \vp {\varphi}
\def \Om {\Omega}

\def \ov {\over}
\def \s{\sigma}
\def \pa{\partial}
\def \ta{\tau}

\def \ha {{1 \over 2}}

\def \la{\label}

\def \k {\kappa}
\def\foot{\footnote}

\def \const {{\rm const}}
\def \NN {{\rm N}} 
\def \J {{\cal J}}
\def \L {\Lambda}
\def\rr {{\rm r}}
\def \st {{\rm t}} 
\def \lle {{\ell}}
\def \sa {\sum_{a=1}^2} 

\newcommand{\rf}[1]{(\ref{#1})}

\renewcommand{\thefootnote}{\fnsymbol{footnote}}

\def\appendix#1{
  \addtocounter{section}{1}
  \setcounter{equation}{0}
  \renewcommand{\thesection}{\Alph{section}}
  \section*{Appendix \thesection\protect\indent \parbox[t]{11.15cm}
  {#1} }
  \addcontentsline{toc}{section}{Appendix \thesection\ \ \ #1}
  }

\newcommand{\eq}[1]{(\ref{#1})}

\def\be{\begin{equation}}
\def\ee{\end{equation}}

\def \ci {\cite}

\def \foot {\footnote}
\def \bi{\bibitem}
\def \tr {{\rm tr}}
\def \ha {{1 \over 2}}
\def \td {\tilde}

\def \ci{\cite}
\def \N {{\cal N}}
 
\def \const {{\rm const}}

\def \ss {\sum_{i=1}^3 }

\def \t {\tau} 
\def\S{{\cal S} }

\def \nn {\nu}
\def \XX {{\rm X}}
\def \Om {\Omega}
\def \vom {{\bar \omega}}
\def \Y{{\rm Y}} 
\def \zz {{\rm z}}
\def \rL {{L}}

\def \n {\nu} 
\begin{document}


\null\vskip-24pt \hfill AEI-2003-091
\vskip-1pt
\hfill
\vskip-1pt
\hfill {\tt hep-th/0311004}
\vskip0.2truecm
\begin{center}
\vskip 0.2truecm {\Large\bf
Spinning strings in $AdS_5\times S^5$:
\vskip 0.2truecm
new  integrable system relations
}
\vskip 1.5truecm
{\bf G. Arutyunov$^{1,}$\footnote{On leave of absence from Steklov Mathematical
 Institute,
Gubkin str.8,117966, Moscow, Russia},
J. Russo$^{3}$
and A.A. Tseytlin$^{2,}$\footnote{Also at Imperial College London
and  Lebedev
 Institute, Moscow.}\\
\vskip 0.4truecm
$^{1}$
{\it Max-Planck-Institut f\"ur Gravitationsphysik,
Albert-Einstein-Institut, \\
Am M\"uhlenberg 1, D-14476 Golm, Germany}\\
\vskip .2truecm
$^{2}$ {\it Department of Physics,
The Ohio State University,\\
Columbus, OH 43210-1106, USA}\\
\vskip .2truecm
$^{3}$ {\it  
Instituci\' o Catalana de Recerca i Estudis Avan\c{c}ats (ICREA),\\
Departament ECM,
Facultat de F\'\i sica, Universitat de Barcelona, 
 Spain}
 }
\end{center}
\vskip 0.5truecm
\vskip 0.2truecm \noindent\centerline{\bf Abstract}
\vskip .2truecm
A general class of rotating closed string solutions in $AdS_5 \times S^5$ 
is shown to be  described by a Neumann-Rosochatius 
one-dimensional integrable system. The latter represents an 
oscillator on a sphere or a hyperboloid 
with an additional ``centrifugal'' potential.
We expect that the reduction of the  $AdS_5 \times S^5$
sigma model to the  Neumann-Rosochatius system should 
have further generalizations and should be  
useful for  uncovering  new relations between integrable  structures on 
two sides of the  AdS/CFT duality. 
We find, in particular,
 new   circular rotating string solutions with
two $AdS_5$ and three $S^5$  spins. 
As in other recently discussed examples, 
the  leading  large-spin correction to  the classical energy 
turns out to be proportional to 
the square of the string tension  or 
the 't~Hooft coupling $\lambda$,
suggesting that it can be matched  onto the 
 one-loop anomalous dimensions
of the corresponding ``long'' operators on the SYM side
of the AdS/CFT duality.

\newpage

\renewcommand{\thefootnote}{\arabic{footnote}}
\setcounter{footnote}{0}

\section{Introduction and summary}

Integrability  of the spin chain 
Hamiltonian representing the planar 
one-loop dilatation operator of $\N=4$ Super Yang-Mills theory 
\cite{mz1}--\ci{bes3}
has recently made possible, following a  proposal  in \ci{ft2,ft3}, 
a number of remarkable and striking tests
of AdS/CFT duality \cite{mz2}--\cite{mz3}. This  generalizes 
the  near-BPS correspondence of \ci{bmn} to non-BPS cases.

The AdS/CFT correspondence predicts that the energy of 
a given physical string state 
(in global $AdS_5$  coordinates)
should match  the scaling
dimension of the corresponding operator in gauge theory.
While the full energy spectrum of the 
quantum string in $AdS_5\times S^5$  is hard to determine, 
 some of its  parts can be probed by considering
the semiclassical string configurations \cite{gkp,ft1}. 
In  certain cases with  large quantum numbers 
(like  
 angular momenta $J_i$ in $S^5$), 
 one finds that the  energy 
of the string solution is given by its  classical expression, i.e.  
 quantum sigma model corrections appear to be suppressed 
\cite{ft3}.

On the gauge theory side, the (one-loop) 
scaling dimensions of 
gauge-invariant composite operators can be found by 
solving the eigenvalue problem for the Hamiltonian of an 
associated spin chain. This is achieved by 
means of algebraic  Bethe ansatz techniques.
In general,  the Bethe ansatz leads to  a complicated system of algebraic
equations. However, 
in the thermodynamic limit (of large
quantum numbers or ``long'' operators) 
the algebraic equations turn into 
integral ones and 
 with some natural assumptions about the density distribution of
Bethe roots the  explicit solutions can be found.
Remarkably,  the Bethe solutions obtained in the 
thermodynamic limit turn out to be  related
to semiclassical string configurations in a precise way.

In general, one can classify strings  moving on $S^5$
with   three ``R-charges'' (SO(6) spins) 
defining the  highest weight state  $(J_1,J_2,J_3)$
of an   SO(6) representation.
For a simpler case of two non-vanishing spins  
$(J_1,J_2)$ the string evolution equations are solved in terms 
of elliptic functions; the corresponding 
string configurations can have folded
\cite{ft4}
or circular \cite{ft2,afrt} profiles, giving rise to two different 
expressions for the space-time energy.
On the gauge theory side,  the relevant Bethe solutions and the
associated scaling dimensions have been found in  \cite{mz2,bfst},
and shown to agree with their string counterparts for 
both   folded \cite{ft4,mz2,bfst} and  
circular \cite{mz2,afrt,bfst} type configurations.
Other surprising examples of a perfect agreement between  string 
energies and  scaling dimensions of gauge theory operators 
include \cite{mz3} a simple circular string 
solution   with three 
spins \cite{ft2} and a pulsating string solution \cite{mina}. 

Even more remarkably, in the  recent work \cite{AS}
the entire Bethe resolvent (corresponding either to the circular 
or to the folded string  type thermodynamic density distributions) 
was  reproduced 
from the classical string sigma model.
This agreement goes beyond comparing just the string energies
with the scaling dimensions: it involves matching  the {\it infinite} 
towers of commuting conserved charges
on the gauge and string sides of the AdS/CFT correspondence.
In fact,  the Bethe resolvent  is nothing else but 
  a generating function 
of local conserved  commuting  charges in string theory 
properly restricted 
to the  leading (${\cal O}(\l)$ or ``one-loop'')  level!

\medskip

 The  matching of  higher local commuting 
string charges \cite{AS}\foot{The integrability of
the O(n) invariant  sigma models was
 discussed, e.g.,   in  \cite{lup}-\ci{ogi}.
 Classical solutions for 
 strings in constant curvature  spaces 
were  studied  in \ci{BN}-\ci{ves} and refs. there.
More recent AdS/CFT motivated 
 discussions concerning integrability, higher
  local and non-local 
 charges  and  Yangian structures of related sigma models
are in  \cite{Mandal}-\cite{Alday}.} 
and  recent advances in study of integrability of 
the dilatation operator at higher 
loops in  $\N=4$ Super Yang-Mills
theory \cite{bes3,Beisert:2003ys} (and in its $S^3$ reduced 
matrix model version \cite{KP})
 provide  strong support that the
 same integrable structure should be underlying 
  the 
 two sides of the duality.

Still, our understanding of the gauge/string 
duality, even in the 
``semiclassical'' (large quantum number) 
 sector of states, is far from complete. More 
detailed analysis of different physical configurations 
in both gauge and string theories is required to  elucidate
how the duality works.
While recent papers \ci{bfst,AS,mz3,Dolan} 
 shed some light on how  
the integrability of the \adss string theory is
 related to that of the 
planar SYM  theory,  many details are  missing. 


In view of the  general problem of establishing
 correspondence between
various  integrable subsectors
of string and gauge theories 
it is of interest to obtain a  systematic picture of  
reductions  of the two-dimensional   integrable 
O(4,2)$\times$O(6) sigma model
describing propagation of the classical  string in the 
\adss\ space-time  to  various one-dimensional integrable models.
In \cite{afrt} we have shown that for a  natural rotating string 
 ansatz the \adss string sigma model 
reduces  to an  integrable  Neumann model 
\cite{Neumann} describing an oscillator on a 2-sphere.


The aim of the present  paper is to make  further progress
 in this direction.
We will consider a more general integrable subsector in string theory 
which arises from a  rotating string  ansatz extending 
 the one in 
\ci{afrt}.
In this case the 2-d sigma model reduces to  
the Neumann-Rosochatius (NR) \cite{gag} integrable system
describing a particle on a sphere in
the  $\sum_i (w^2_i r^2_i + v^2_i r^{-2}_i)$
potential (in the previous  case \ci{afrt} we had $v_i=0$). 
While, as  in \ci{afrt}, the 
 general solutions of this system are given  by 
theta-functions on a genus 2 hyperelliptic curve
its new feature is the existence of a very simple new class 
of solutions corresponding to circular strings 
with constant radii $r_i(\sigma)=$const. 
These solutions generalize the ones of \ci{ft2} (which 
had two equal  spins)
to the case when all 2+3 \adss spins may be different.
The corresponding energy  has a very simple dependence on the spins 
and winding numbers. 
Understanding its  SYM scaling dimension  counterpart
 should help, in particular,  to clarify the issue of 
 how the winding  numbers of  circular strings 
  are reflected in the Bethe root distributions (cf. \ci{mz2,afrt,mz3}).

\bigskip

Let us now  summarize the  contents of the paper.
In section 2.1 we shall present the generalized rotating string ansatz 
for a closed string fixed at the origin of $AdS_5$ and 
rotating in 3 orthogonal planes in $S^5$ and explain 
the reduction of the O(6) invariant sigma model to the NR system
for the 3 radial directions of the string. 
In section 2.2 we will list  the corresponding integrals 
of motion and the 
Virasoro constraints  allowing one to express the $AdS_5$ energy as a
function of the three  $S^5$ spins.
In section 2.3 we shall mention that a ``2d-dual'' version  of the 
rotating string ansatz (with roles of $\t$ and $\s$ interchanged)
describes a general pulsating string solution 
with radii oscillating in time which is thus  also described 
by an NR integrable model
(some  special cases of pulsating solutions were previously 
discussed in \ci{vesa,gkp,mina,Khan,mz3}). 
In section 2.4 we shall clarify how the integrability 
of the NR system  follows from  its relation to the 
integrable   O(6) sigma model by deriving its 
   Lax representation.
We shall also explain  how higher commuting charges 
 can be computed from 
the \sm\ monodromy function.

In section 3 we shall study a
 very simple special class of NR solutions on $S^5$ 
which has a similarity with rotating string solutions 
in flat space and generalizes the circular 2-spin and 3-spin 
rotating string 
solutions in \ci{ft2}. As will be shown in section 3.2, the 
 corresponding energy  has a regular large-spin 
expansion in ${\l \ov J^2}$.
 In section 3.3 we shall find the spectrum of  quadratic 
 fluctuations near these circular solutions 
 (extending and simplifying the discussion in \ci{ft3} for the 
 special solutions of \ci{ft2}). We shall determine the 
 stability conditions and mention some  
 straightforward  applications. 
 

  In section 4 we shall study 
 more general  solutions
  of the NR  system  with a non-trivial 
  dependence on the world-sheet coordinate $\s$. We 
  shall consider, in particular,  a 
two-spin solution which is expressed in terms of the 
elliptic functions. The resulting system of equations relating 
energy and two spins turns out to be more involved that in the previously
discussed elliptic (sine-Gordon) limit of the Neumann model  
 \ci{ft4,afrt,bfst}, but we expect that  it might
  be possible to directly match an appropriate ``one-loop''  limit 
  of this system  onto the corresponding Bethe ansatz 
 equations on the SYM side 
(as was done in the Neumann model case in \ci{mz2,bfst}).

Finally, in section 5 we  shall generalize the discussion of
 sections 2 and 3 to the case when the string can 
 rotate in both $AdS_5$ 
and $S^5$. Here we get a 
 combination of the 
two NR systems (an $AdS_2$ and $S^2$  one)
coupled   by the Virasoro constraints. 
We again consider the simplest  solution with constant radii 
parametrized by 2+3 spins $(S_a,J_i)$  and 2+3 winding numbers.
If the string rotates only in $AdS_5$ the corresponding energy 
does not have a regular large-spin expansion  (section 5.1), 
but it does  if there is at  least one large spin in 
$S^5$  (section 5.2). 
For example, the simplest $(S,J)$ string solution which is
 a circle in both $AdS_5$
and $S^5$ is stable, and 
 it  should be possible  to  match the leading large $J$ correction 
 to its energy   with  a
  particular anomalous dimension 
on the SYM side by identifying 
the corresponding distribution of Bethe roots  
in the  associated  XXX$_{-1/2}$  spin chain \ci{bes2} 
(as was done for other folded and circular $(S,J)$  string 
solutions in \ci{bfst}).

\setcounter{equation}{0}

\section{Reduction of  O(6) sigma-model to the \\ 
Neumann-Rosochatius system}

\subsection{Generalized rotating string ansatz
}
Here we shall generalize the rotation ansatz 
in \ci{afrt} which allowed us to reduce the classical 
string \sm\ equations to those of a 1-d integrable model. 
That will lead to new interesting  simple
classes of rotating string solutions.

Let us 
consider the bosonic part of the classical closed 
string propagating in the  
$AdS_5\times S^5$ space-time. The world-sheet action 
in the  conformal gauge is  
\be
I= - { \sql  \ov 4\pi }
\int d\tau d\sigma  \ \big[ G^{(AdS_5)}_{mn}(x)
\del_a x^m \del^a  x^n\ + \    G^{(S^5)}_{pq}(y)  \del_a y^p
\del^a y^q \big] \ , \ \ \ \ \ \ \ \ \sql \equiv  { R^2 \ov
\aa} \ .  \la{A}
\ee
It is convenient to represent \eq{A}
as an action for the O(6)$\times $SO(4,2) 
sigma-model (we follow the notation of \ci{ft2}) 
\be
I={ \sql  \ov 2\pi }\int d\tau d\sigma (L_S+L_{AdS})\ 
, \la{Lsp} \ee
where
\bea
\la{SL}
L_S&=&-\frac{1}{2}\pa_a X_M\pa^a X_M+\frac{1}{2}\Lambda 
(X_MX_M-1)\, , \\
L_{AdS}&=&-\frac{1}{2}\eta_{MN}\pa_a Y_M\pa^a Y_N
-  \frac{1}{2}
\td \Lambda (\eta_{MN}Y_MY_N+1)\, .
\la{SSL}
\eea
Here  $X_M$, $M=1,\ldots , 6$  and $Y_M$, 
 $M=0,\ldots , 5$ are  the  
the embedding coordinates of 
${R}^6$ 
with the Euclidean metric in $L_S$ and 
 with  $\eta_{MN}=(-1,+1,+1,+1,+1,-1)$ 
 in $L_{AdS}$ respectively.
 $\Lambda$  and $\td \Lambda$ are the Lagrange multipliers.
The action \eq{Lsp} is  to be supplemented 
 with the usual conformal gauge  constraints.
The embedding coordinates of \adss can be parametrized in terms of
angles  of $AdS_5$ and $S^5$ as in \ci{ft2,afrt} 
\be\la{relx}
 X_1 + i X_2 = \sin   \g \ \cos
\psi \ e^{ i \vp_1} \ , \ \ \ \
X_3 + i X_4 =  \sin   \g \ \sin \psi \
e^{ i \vp_2} \ , \  \ \ \ \ \ 
X_5 + i X_6 = \cos  \g \ e^{ i \vp_3} \ ,
\ee
\be \la{rell}
Y_1+ iY_2 = \sinh \r \ \sin \theta \  e^{i \phi_1}\ , \ \ \ \ \
Y_3 + i Y_4  = \sinh \r \ \cos \theta \  e^{i  \phi_2}\  , \ \ \ \
Y_5 + i Y_0 = \cosh \r \  e^{i t } \ . \ee
In this section we  will 
be discussing the case when the 
string is located at the center of $AdS_5$ 
and  rotating in $S^5$, i.e.
is trivially embedded in $AdS_5$ as $Y_5+iY_0=e^{it},$
 with the global time of $AdS_5$ being $t=\k \tau$ and 
with  $Y_1,...,Y_4=0$.

The $S^5$ metric 
has three commuting translational
isometries in $\vp_i$ in \rf{relx}
which give rise to three global 
commuting integrals of motion (spins) $J_i$. Since we are 
interested in a periodic motion with  $J_i\not=0$ 
it is natural to choose the following {\it ansatz} for $X_M$:
$$
\XX_1\equiv X_1+iX_2=z_1(\sigma)\ e^{iw_1\tau}\, , ~~~\ \ \ \ \ \ \ 
\XX_2\equiv X_3+iX_4=z_2(\sigma)\ e^{iw_2\tau}\, , ~~~ $$
\be
\la{emb}
\XX_3\equiv X_5+iX_6=z_3(\sigma)\ e^{iw_3\tau}\, .
\ee
In contrast to our earlier work  \ci{afrt} here we shall
 not assume that $z_i$ are real, i.e. in general 
\be \la{hah}
z_k =\  r_k(\s)\   e^{ i \a_k(\s)}  \ , \ \ \ \ \ \ \ \ \ \ k=1,2,3\ . 
\ee
In order to find the relevant closed string  solutions
 we need also to impose 
the periodicity conditions  on  $X_M$ or  $z_i$:
\be\la{pep}  r_i (\s + 2 \pi) = r_i (\s ) \ , 
 \ \ \ \ \ \ \ \ 
\a_i(\s+ 2 \pi ) =  \a_i + 2 \pi  m_i  \ , \   
  \ \ \ \ \ m_i =
0,\pm 1,\pm 2, ... \ . \ee 
Thus   $r_k$ are   real periodic functions of $\s$, 
while real phases  $\a_k$ are 
 periodic only up to $2\pi m_k$ shift.
 
 Comparing \rf{emb} to \rf{relx} we conclude that 
 for this  general ``complex'' ansatz the angles 
 $\vp_i$ depend on both $\tau$ and $\sigma$, 
 \be \vp_i = w_i \tau + \a_i (\s)\ . \ee
 The integers $m_i$ 
 that will label different solutions 
 thus play  the role of ``winding numbers'' in the 
 linear isometry
 directions $\vp_i$. 

As a consequence of $X_M^2=1$, 
$r_k$ must  lie on a two-sphere:  
\be 
\sum^3_{i=1} r_i^2=1 \  . \ee
The space-time energy $E$ of the string
 (related to the  generator  of a
compact SO(2) ``05'' subgroup of SO(4,2)) here  is simply
\bea \la{enn}
E=\sqrt{\lambda}\ \kappa\equiv \sqrt{\lambda}\ \E \, .
\eea
The  spins $J_1=J_{12}, \ J_2=J_{34}, \ J_3=J_{56}$ 
forming  a  Cartan subalgebra of  SO(6)  are
\be
\la{spins}
J_i=\sqrt{\lambda}\ w_i
\int_0^{2\pi}\frac{d\sigma}{2\pi}\ 
 r_i^2(\sigma)\equiv\sqrt{\lambda}\ {\cal J}_i \, , 
\ee
and thus satisfy 
\be\la{genn}
\ss { \J_i \ov w_i} = 1 \ . \ee
As discussed  in \ci{ft2}, to have a consistent 
semiclassical string state interpretation of these 
configurations one should  look for solutions  for which 
all other components of the SO(6) angular momentum tensor 
$J_{MN}$ vanish.
This is  automatically   the case  if all $w_i$ are 
different \ci{afrt}, but are to be checked in other cases.

The  
Virasoro constraints that need to be imposed on a 
sigma model solution of \rf{SL}  are 
(dot and prime are derivatives over $\tau$ and $\sigma$)
\be \la{cv}
\kappa^2=\dot{X}_M\dot{X}_M+ X_M ' X_M '
=\sum_{i=1}^3 ( r'^2_i    + r^2_i \a'^2_i   + w_i^2  r^2_i )  
  \, , 
\ee
\be \la{cvv}
0=\dot{X}_M{X'}_M= 2\sum_{i=1}^3w_i r^2_i \a'_i    \ . 
\ee

\subsection{Integrals of motion  and constraints
}
In general, starting with 
\be \la{ah}
\XX_i(\t,\s) = r_i(\t,\s) e^{i\vp_i(\t,\s)} \ee
we get from \rf{SL}  the Lagrangian 
\be
\label{lah}
L_S=\frac{1}{2}\sum^3_{i=1} \big[  {\dot r}_i^2
- r'^2_i + r^2_i ( {\dot \vp}_i^2 
- \vp'^2_i  ) \big] +  \frac{1}{2}
\Lambda(\sum^3_{i=1}  r^2_i-1) \, .
\ee
One can easily check that the ansatz 
\be \la{gig}
r_i = r_i (\s) \ , \ \ \ \ \ \ \ \  \
\vp_i = w_i \t + \a_i (\s) 
\ee
is indeed consistent with the equations of motion.

Substituting the ansatz  (\ref{gig}) or \rf{emb}
into the SO(6) Lagrangian   \eq{SL}
 we get the following effective  1-d ``mechanical''  system
for a particle on a  5-d sphere 
(we change the sign of $L$ since now  $\s$ plays the role 
of 1-d time) 
\be
\label{L}
L=\frac{1}{2}\sum^3_{i=1} ( z'_i z'^*_i -  w_i^2  z_i
z^*_i)- \frac{1}{2}
\Lambda(\sum^3_{i=1}  z_i z^*_i-1) \, .
\ee
If we set  $z_k = x_k + i  x_{k+3}$, this is recognized 
as a special case of the standard integrable 
$n=6$  Neumann model
(harmonic oscillator on a 5-sphere)  where three
of the six frequencies are equal to the other three. 
This relation  implies integrability of \rf{L} 
 model, i.e.  determines integrals of motion. 

Equivalently, in the  ``planar'' coordinates \rf{hah} 
we get from \rf{lah} 
\be
\label{La}
L=\frac{1}{2}\sum^3_{i=1} ( r'^2_i  + r^2_i \a_i'^2  -  w_i^2  r^2_i)- 
\frac{1}{2}
\Lambda(\sum^3_{i=1}  r^2_i-1) \, .
\ee
Equations for the angles $\a_i$ can be  integrated once
\be\la{ank}
\a'_i = { v_i \ov r^2_i } \ ,\ \ \ \ \ \ v_i= {\rm const} \ , 
\ee
where $v_i$ are three  integrals of motion. 
Substituting $\a'_i$ 
  back into the action we get the following 
  effective  Lagrangian for the 3 real radial coordinates
\be
\label{Laa}
L=\frac{1}{2}\sum^3_{i=1} \big(  r'^2_i   -   w_i^2  r^2_i -
   { v_i^2 \ov
r_i^2}
\big)-  \frac{1}{2}
\Lambda(\sum^3_{i=1}  r^2_i-1) \,  . 
\ee
When the new integration constants $v_i$ vanish, i.e.
$\a_i$ are constant, 
we go back to the previously studied \ci{afrt} 
 example 
 of the $n=3$ Neumann model.
 For non-zero $v_i$ the Lagrangian \rf{Laa}
 describes the so called {\it  Neumann-Rosochatius} (NR) 
 integrable system (see, e.g., \ci{gag}).
  Its integrability follows  already from 
 the fact that it is  a special case
  of the 6-dimensional Neumann system. 
 

Finding the integrals  of the ``radial'' 
system \rf{Laa}  is straightforward using the 
relation to 
the  Neumann model: the 
 $n=6$  Neumann system with coordinates $x_M$ 
  has, in general, the following 
six   integrals of motion:
\bea \la{inti} 
F_M=x_M^2+\sum^6_{M\neq N}\frac{(x_M x'_N-x_N x'_M)^2}
{w_M^2-w_N^2} \,  , \ \ \ \ \ \ \ \ \ \ 
\sum^6_{M=1}  F_M=1 \ .
\eea
However, in our case there are equalities between frequencies 
($w_1=w_4, w_2=w_5, w_3=w_6$) 
so one should be careful to avoid singularities.
The integrals of the Neumann-Rosochatius model are  obtained as
the following combinations $I_i= F_i + F_{i+3}$ ($i=1,2,3$)
 in which singular terms cancel. 
Explicitly, we find (using \rf{ank}) 
\be\la{nti}
I_i =
r^2_i + \sum^3_{j\not=i} { 1 \ov w^2_i - w^2_j} 
\bigg[ (r_i r'_j - r_j r'_i)^2  + { v^2_i \ov r^2_i} r^2_j 
+ { v^2_j \ov r^2_j} r^2_i  \bigg] \ , 
\ \ \ \ \ \  \ss I_i=1 \ . 
\ee
This gives us two  independent  integrals of motion (which we shall denote
 $b_a$) 
in 
addition to the  three  other 
integrals ($v_i$) we found  already.

The constraints  \rf{cv},\rf{cvv} can be written as :
\be \la{cve}
\kappa^2=\sum_{i=1}^3( r'^2_i      + w_i^2 r^2_i  +  
 { v_i^2 \ov r^2_i} )    \ , 
\ee
\be \la{cvve}
\ \sum_{i=1}^3 w_i v_i  =0   \ .
\ee
As a consequence of \rf{cvve}
only   two  of  the three integrals of motion   $v_i$ are 
 independent of $w_i$. 


\bigskip

As discussed  in
\ci{afrt}, the periodicity condition in \rf{pep} on $r_i$ 
  implies  that the integrals of motion $b_a$
 can be traded for two integers $n_a$  labeling different 
  types of solutions. 
Imposing the periodicity condition in \rf{pep} on $\a_i$  
gives, in view of \rf{ank}, the following constraint:
\be\la{okl}
v_i \int^{2\pi}_0 { d\s \ov r^2_i(\s)} = 2 \pi m_i  \ . \ee
It implies that $v_i$ should be expressible in terms of the 
integers $m_i$, frequencies $w_i$ and the ``radial'' integrals $b_a$
or $n_a$.\foot{Note that since the integral in \rf{okl}
is of a positive function
$m_i=0$ implies $v_i=0$.}
The moduli  space of solutions 
  will thus be parametrized  by 
  $(w_1,w_2,w_3; n_1,n_2; m_1,m_2,m_3)$.
  The constraint \rf{cvve}  will give one relation between these 
3+2+3 parameters. As a consequence, trading $w_i$ for the 
angular momenta, the energy of the solutions 
as determined by \rf{enn},\rf{cve} will be 
a 
  function of the SO(6) spins and the 
 ``topological'' numbers $n_a$ and $m_i$
\be \la{ennp} 
\E=\E(\J_i; n_a,  m_i)\ , \ \ \ \ \ \ \ 
E= \sql\ \E ( { J_i \ov \sql}; n_a,  m_i)  \ . \ee 
The constraint \rf{cvve} will provide one  additional relation 
between $J_i$ and $n_a,m_i$. 
 
 \bigskip
 
In the following sections of   this paper   we shall consider several special solutions 
 of the above system \rf{La}. 
 We shall  start in section 3 with a discussion of the 
  simplest possible solution with constant $r_i$ 
  (for which $n_a=0$) and 
 which  represent   an interesting new class  of  circular
  3-spin solutions generalizing  the circular  solution of 
   \ci{ft2}.
   
    \bigskip
    
\subsection{``2d-dual'' NR system for pulsating solutions}

It  is of interest to consider a ``2d-dual'' version of the
 rotation ansatz
\rf{emb},\rf{hah} where $\tau$ and $\sigma$ 
are interchanged (but still  keeping  the $AdS_5$ time as 
$t=\k \tau$), i.e. 
\be\la{tat}
\XX_i = z_i(\tau)\ e^{i m_i \s} = 
r_i(\tau) e^{ i \a_i (\t) + i m_i \s} 
  \ , \ \ \ \ \ \  \ \ \ \ \ \ \ \ss r^2_i(\t) =1 \ . \ee
 In this case the radial directions depend on $\t$ instead of $\s$ 
  and the ``frequencies''  $m_i$ must take integer values 
 in order  to satisfy the closed string periodicity 
 condition. In general, in order to have the zero 
  non-Cartan components of the 
 O(6) angular momentum tensor one is to assume that $m_i \not=m_j$.

  This ansatz  describes an ``oscillating'' or ``pulsating''
  $S^5$  string configurations, 
   special cases of which (with motion in both $AdS_5$ and
   $S^5$) 
   were discussed previously in \ci{vesa,gkp,mina,Khan,mz3}.
   
   Since the sigma model  Lagrangian \rf{SL} is 
   formally invariant under $\s \leftrightarrow \t$, the resulting  
1-d effective Lagrangian will have essentially 
the same form as \rf{L},\rf{La} (here we do not invert the sign of
the Lagrangian) 
 \be
\label{Le}
L=\frac{1}{2}\sum^3_{i=1} ( \dot z_i \dot z^*_i -  m_i^2  z_i
z^*_i) +  \frac{1}{2} \Lambda(\sum^3_{i=1}  z_i z^*_i-1) \, .
\ee
Solving for $\dot \a_i$ as in \rf{ank} we get  
$r^2_i \dot \a_i = \J_i$=const, where  the counterparts of 
the integration constants $v_i$  are, in fact, 
 the angular momenta in \rf{spins}.
 Then  we end up with the following analogue of 
 \rf{Laa}
  \be
\label{Lae}
L=\frac{1}{2}\sum^3_{i=1} \big(  \dot r^2_i   -   m_i^2  r^2_i -
   { \J_i^2 \ov r_i^2}
\big)  +   \frac{1}{2} \Lambda(\sum^3_{i=1}  r^2_i-1) \,  . 
\ee
Thus pulsating solutions (carrying also 3 spins $\J_i$) 
are again  described by  a special 
Neumann-Rosochatius integrable system. 

Since (the $S^5$ part of) the  corresponding conformal
gauge constraints are also $\t \leftrightarrow \s$ symmetric,  
they take form similar  to \rf{cv},\rf{cvv} or \rf{cve},\rf{cvve}
\be \la{cvek}
\kappa^2=\sum_{i=1}^3( \dot r^2_i  + 
m_i^2 r^2_i  +  { \J_i^2 \ov r^2_i} )    \ , 
\ee
\be \la{cvvek}
\ \sum_{i=1}^3  m_i \J_i  =0   \ .
\ee
 One may look for   periodic  solutions of the above NR
 system \rf{Lae} subject to the constraint \rf{cvek},
  i.e. having finite 
 1-d energy (equal to $\ha \k^2$). 
In the simplest (``elliptic'') case reducing to a sine-Gordon 
type system we may follow \ci{dhn,vesa,mina} and  introduce, as
  for any periodic solitonic solution,  an 
   oscillation ``level number'' $\NN $.
  This  may be achieved  by   considering a
 semiclassical (WKB) quantization of the action \rf{Lae}. 
 
 Here we  shall not go into detailed study of the resulting 
 pulsating string solutions.
 Let us only mention that 
 a special $r_i=$const solution  of the above system 
 (when, in fact, 
  there is  no oscillation of the  radii)  is essentially the same 
 as the special circular solution with 
 $r_i=$const of the system \rf{Laa} discussed below in  section 3. 
  
  In the case of the $S^5$ pulsating solution in \ci{mina,mz3} 
  the expansion of the energy at large level $\NN\gg 1$ appears to be 
  regular in $\l\ov \NN^2$ (this is not the case for pulsating string in 
  $AdS_5$ \ci{mina}) and, indeed, the leading $\l\ov \NN^2$ term in $E$ 
  can then be matched onto the SYM 
  anomalous dimensions  as was shown in \ci{mz2,mz3}.

 \bigskip 
 
\subsection{Lax representation for the NR
system induced from the O(6) sigma model}

Having in mind further generalizations, it is 
useful  to understand how the integrability of the NR system 
(e.g.,  the Lax representation) follows from the fact that this system 
is embedded into a much more general integrable \ci{lup}
O(6) sigma model.
Here we will clarify this issue and point out some related open 
problems.

We start with describing  the zero-curvature representation for the O(6)
sigma-model in terms $4\times 4$ matrices. Let $\XX_i$ be the 3 complex 
embedding fields \rf{emb}  of the O(6) model. Let us introduce the following 
skew-symmetric
matrix $S$:
\bea
S=\left(\begin{array}{cccc}
0 & \XX_1 & -\XX_2 & \bar{\XX}_3 \\
-\XX_1 & 0 & \XX_3 & \bar{\XX}_2 \\
\XX_2 & -\XX_3 & 0 & \bar{\XX}_1 \\
-\bar{\XX}_3 & -\bar{\XX}_2 & -\bar{\XX}_1 & 0 
\end{array}\right) \, .
\eea
The matrix $S$ is also unitary,  $SS^{\dagger}=1$, 
 provided $\XX_i\bar{\XX}_i=1$. Let us also
introduce the su(4)-valued current $A$ with components
\bea
\label{csm}
A_{\tau}=S\pa_{\tau}S^{\dagger}\, , ~~~~~
A_{\sigma}=S\pa_{\sigma}S^{\dagger}\, , \ \ \ \ \ \ 
A_\pm = \ha ( A_\ta \pm A_\s) \ . 
\eea
This current can be  used to construct the following matrices $U$ and $V$ \cite{lup}: 
\bea
U= \frac{1}{1+\lle} A_- 
-\frac{1}{1-\lle} A_+ \, \ \ \ \ \ \ \ \ \ \ 
V= -\frac{1}{1+\lle} A_- 
-\frac{1}{1-\lle} A_+ \  .
\eea
Here $\lle$ is a  spectral parameter,  and by construction 
$U$ and $V$ have  simple poles at $\lle=\pm 1$. 
They obey the zero-curvature condition
\bea
\label{zc}
\pa_{\tau}U-\pa_{\sigma}V+[U,V]=0\ , 
\eea 
which is a crucial device for demonstrating the 
integrability of the sigma models.
Quite generally,  one can associate to eq.(\ref{zc}) the transition matrix 
${\mbox T}(\sigma,\lle)$ (see,  e.g.,  \cite{FT}) 
defined through the path-ordered exponent:
\bea
{\mbox T}(\sigma,\lle)=\mbox{P}\exp{\int_0^{\sigma}U(\sigma',\lle){\rm d}\sigma'}
\ , \eea
and show that the trace of the  monodromy matrix (the parallel transport along the period of the 
zero-curvature connection)
\bea
\label{monodromy}
{\cal Q}(\lle)=\mbox{Tr}\ {\mbox T}(2\pi,\lle)
\eea 
generates (when expanded as ${\cal Q}= \sum^{\infty}_{n=0} {\cal Q}_n \lle^n$) an  infinite tower of 
commuting integrals of motion.\footnote{
Derivation of the Poisson algebra satisfied by matrix elements of the transition matrix
and the proof of commutativity of the integrals generated by the monodromy matrix
can be found in \cite{M1}. }

Consider now the generalized rotation (or ``Neumann'')
  ansatz for the sigma model variables $\XX_i$
in \rf{emb}, i.e. 
\bea
\label{genNA}
\XX_i=z_i(\sigma)e^{iw_i\tau}\, , \ \ \ \ \ \ \ \ \ss |z_i|^2 =1 \ . 
\eea 
Remarkably, the current (\ref{csm}) evaluated on $\XX_i$  of the form
(\ref{genNA}) admits the following factorization
\bea
A_{\tau}=Q(\tau){\cal A}_{\tau}Q^{\dagger}(\tau)\, , ~~~~
A_{\sigma}=Q(\tau){\cal A}_{\sigma}Q^{\dagger}(\tau)\, .
\eea
Here $Q(\tau)$ is the diagonal matrix 
$$
Q(\tau)=\mbox{diag}\Big(
e^{-iw_3\tau}  ,\ e^{-iw_2\tau}  , \ e^{-iw_1\tau}  ,\ 
e^{-i(w_1+w_2+w_3)\tau} \Big) \, ,
$$
while the matrices ${\cal A}_{\tau}$ and ${\cal A}_{\sigma}$
are independent of $\tau$ and given by 
 \bea
\nonumber{\footnotesize
{\cal A}_{\tau}=i
\left(
\begin{array}{rrrr}
2w_3z_3z^*_3-w_iz_iz^*_i & (w_2+w_3)z_2z^*_3 & (w_1+w_3)z_1z^*_3 & (w_1-w_2)z_1z_2 \\
(w_2+w_3)z_2^*z_3 & 2w_2z_2z_2^*-w_iz_iz^*_i  & (w_1+w_2)z_1z^*_2 &  -(w_1-w_3)z_1z_3 \\
(w_1+w_3)z^*_1z_3 & (w_1+w_2)z^*_1z_2 & 2w_1z_1z^*_1-w_iz_iz^*_i & (w_2-w_3)z_2z_3 \\
(w_1-w_2)z^*_1z^*_2 & -(w_1-w_3)z^*_1z^*_3 & (w_2-w_3)z^*_2z^*_3 & w_iz_iz^*_i
\end{array}\right) ,}
\eea 
and
\bea\nonumber{\footnotesize
{\cal A}_{\sigma}=
\left(\begin{array}{cccc}
z_1{z_1'}^*+z_2{z_2'}^*+z_3^*z_3' & z_3^*z_2'-\bar{z_3'}^*z_2 
& z_3^*z_1'-{z_3'}^*z_1 &  z_2z_1'-z_2'z_1  \\
z_2^*z_3'-{z_2'}^*z_3 & z_1{z_1'}^*+z_3{z_3'}^*+
z_2^*z_2' & z_2^*z_1'-{z_2'}^*z_1 &  z_1z_3'-z_1'z_3 \\
z_1^*z_3'-{z_1'}^*z_3 &  z_1^*z_2'-{z_1'}^*z_2   &  
 z_2{z_2'}^*+ z_3{z_3'}^*+z_1^*z_1' &  z_3z_2'-z_3'z_2 \\
z_1^*{z_2'}^*-{z_1'}^*{z_2}^* &  {z_3}^*{z_1'}^*-{z_3'}^*{z_1}^* & 
z_2^*{z_3'}^*-{z_2'}^*{z_3}^*  & z_i'{z_i}^* 
\end{array}\right) .}
\eea
As the consequence,  one finds
\bea
U=Q(\tau){\cal U}(\s) Q^{\dagger}(\tau)\, , ~~~~
V=Q(\tau){\cal V}(\s) Q^{\dagger}(\tau)\, ,
\eea
where ${\cal U}$ and ${\cal V}$ depend only on $\s$. The zero-curvature condition
(\ref{zc}) reduces to
\bea
\pa_{\sigma}{\cal V}=\ [Q^{\dagger}\pa_{\tau}Q-{\cal V},\ {\cal U}]\, .
\eea
Next,  we note that the diagonal matrix $Q^{\dagger}\pa_{\tau}Q$ is $\sigma$-independent 
and, therefore, one can introduce the following $L$ and $M$-operators:
\bea
\label{Ls}
L\equiv {\cal V}-Q^{\dagger}\pa_{\tau}Q \, , \ \ \ \   ~~~~M\equiv -{\cal U} \ , 
\eea
which furnish the Lax representation for the NR system:
\bea
\label{Leq}
\pa_{\sigma}L=[L,M] \, .
\eea
This is a {\it new} Lax representation for the NR system,
the previously known examples include the formulation of the Lax equations 
in terms of $3\times 3$ \cite{AvM} or $2\times 2$ \cite{AT} matrices.   
Thus, the O(6) sigma model indices the Lax pair for the NR system 
in terms of traceless antihermitian $4\times 4$ matrices.
An interesting open problem is to 
construct 
the classical $r$-matrix corresponding to the Lax system (\ref{Ls}), (\ref{Leq}).

As was discussed in the previous subsection, the NR system has 
the ($\sigma$-independent) integrals precisely in number
which is required for its
Liouville integrability. 
 Regarding now $\sigma$ as 
a  (periodic) time variable,  the integrals of motion of the NR system 
can be  constructed, e.g.,  as $F_n=\mbox{Tr}\ L^n$. 
However, being embedded into the more general 
two-dimensional integrable system it inherits 
 an {\it infinite}
number of  conserved (i.e. $\tau$-independent!) integrals of motion.
One possible way to exhibit this infinite commuting family is to compute
the monodromy (\ref{monodromy}) for the Neumann connection $U(\sigma,\lle)$.
In general,  this is a difficult problem, but  it can be simplified by 
considering the special (simplest) solutions of the NR system.

A significant  simplification of the Lax pair 
occurs if we restrict ourselves to the two-spin solutions, which are obtained by 
setting  $\XX_3=0$. In this case we have effectively the SO(4) sigma model
that is isomorphic to two copies of SU(2) models. Indeed, one can show that by  
a similarity transformation the matrices 
${\cal A}_{\tau}$ and ${\cal A}_{\sigma}$
can be brought to the form
\bea
\nonumber {\footnotesize
{\cal A}_{\tau}=i
\left(\begin{array}{cccc}
w_2z_2z_2^*-w_1z_1z_1^*  & (w_1+w_2)z_1z_2^*            &   0  &  0       \\
(w_1+w_2)z_1^*z_2             & w_1z_1z_1^*-w_2z_2z_2^* &    0  &  0     \\
0  & 0 & -w_1z_1z_1^*-w_2z_2z_2^* & (w_1-w_2)z_1z_2 \\
0 &  0 & (w_1-w_2)z_1^*z_2^* &  w_1z_1z_1^*+w_2z_2z_2^*
\end{array}\right)  }
\eea
and 
\bea\nonumber{\footnotesize
{\cal A}_{\sigma}=
\left(\begin{array}{cccc}
z_1{z_1'}^*+z_2^*z_2'  & z_2^*z_1'-{z_2'}^*z_1 &  0  &  0       \\
z_1^*z_2'-{z_1'}^*z_2   & z_2{z_2'}^*+z_1^*z_1' &  0  &  0       \\
0  & 0 &z_1{z_1'}^*+z_2{z_2'}^* & z_2z_1'-z_2'z_1  \\
0 &  0 &z_1^* {z_2'}^*-{z_1'}^*{z_2} & z_1'z_1^*+z_2'z_2^*
\end{array}\right) ,}
\eea
which exhibits factorization into two SU(2) sectors. It is easy to see that the NR 
evolution equations arise already from a single 
SU(2) sector, {e.g.},  from the upper left conner 
of the Lax matrices. Schematically,
the corresponding $L$-operator reads as 
\bea
L=L_0+\frac{L_1}{1-\lle}+\frac{L_{-1}}{1+\lle} \, , 
\eea
were 
$L_0=\mbox{diag}\Big( -iw_2 , -iw_1\Big)$.

Let us recall  that the Neumann model admits two different kinds of 
two-spin solutions 
corresponding to string configurations of the folded or circular type respectively
\cite{afrt}.
For instance,  the two-spin circular type solution can be written in terms of 
the standard Jacobi elliptic functions as follows $(z_3=0)$
\bea
\label{cs}
z_1(\sigma)=\mbox{sn}(a\sigma, \st )\, , ~~~
z_2(\sigma)=\mbox{cn}(a\sigma, \st)\, , ~~~\,\ \ \ \ \ 
a\equiv \sqrt{\frac{w_{12}^2}{\st}}= \frac{2}{\pi}\ellK(\st)\  ,
\eea
where $w_{12}^2= w^2_1 - w^2_2 $ is related to the elliptic modulus 
$\st$ through the closed string 
periodicity condition.\foot{$\ellK$ is the standard complete 
elliptic integral of the first kind.} The modulus $\st$ is related to 
the spins $\cJ_1$ and $\cJ_2$
by a  transcendental equation 
(see \cite{afrt} for details).  
On this particular solution the matrices $U$
and $V$ projected on  the first
SU(2) sector are
\bea
\nonumber {\footnotesize
\hspace{2mm}~U=\frac{1}{1-\lle^2}
\left(\begin{array}{cc}
i\ell(w_1\sn^2 a\sigma-w_2\cn^2 a\sigma)  & 
-a\dn a\sigma-i\ell (w_1+w_2)\sn a\sigma \cn a\sigma             \\
a\dn a\sigma-i\ell(w_1+w_2)\sn a\sigma \cn a\sigma    
     &   -i\ell(w_1\sn^2 a\sigma-w_2\cn^2 a\sigma)
\end{array}\right)   }
\eea
and
\bea
\nonumber {\footnotesize
\hspace{2mm}~V=\frac{1}{1-\lle^2}
\left(\begin{array}{cc}
i(w_1\sn^2 a\sigma-w_2\cn^2 a\sigma)  & 
-a\ell \dn a\sigma-i(w_1+w_2)\sn a\sigma \cn a\sigma             \\
a\ell\dn a\sigma-i(w_1+w_2)\sn a\sigma \cn a\sigma    
     &   -i(w_1\sn^2 a\sigma-w_2\cn^2 a\sigma)
\end{array}\right) \, .  }
\eea
Using these matrices and applying the (recurrent)
abelianization procedure of Zakharov and Shabat \cite{ZS}
one can compute the monodromy function 
 ${\cal Q}(\lle)$ in \rf{monodromy} 
 and, as a  consequence,
the corresponding higher commuting charges.

In a  recent work  \cite{AS} the higher commuting (local)
charges were obtained for both the folded and the circular two-spin  solutions
of the Neumann model and linked with those of the one-loop
 planar ${\cal N}=4$ SYM theory.  
The approach of \cite{AS} was based on finding the form of 
 B\"acklund transformations, which also provides a 
 way of generating  
the commuting conserved charges (see, e.g., \ci{ogi}). 
It would be of interest to understand better a relation  
between the B\"acklund transformations 
and the monodromy approach in our stringy context.

\setcounter{equation}{0}

\section{Special circular solutions:  constant $\L$ case}

A very simple special class of solutions of the system \rf{L} or \rf{La}
which has a similarity with rotating string solutions 
in flat space and generalizes the circular rotating string  solutions in 
\ci{ft2,ft3} has the property that the Lagrange multiplier 
is constant, i.e. $\Lambda=\const$.

\subsection{Constant radii solution} 

Let us 
 start with the Lagrangian 
  \rf{L} in terms of the  complex coordinates
$z_i$. Then  the equations of motion  are 
\be \la{jk} 
z''_i   +  m^2_i z_i = 0  \ , \ \ \ \ \ \ \ \ \ \ \ \ \ \ 
m^2_i \equiv w^2_i + \L \ , \ \ \ \ \ \ \ \ \ss |z_i|^2 =1 \ ,   \ee
\be\la{ppl}
 \L=\ss (|z'_i|^2   -  w^2_i | z_i|^2)\ . \ee
Eq. \rf{jk} 
can be easily integrated if one assumes that $\L=\const$, 
\be \la{uou} 
z_i = a_i e^{i m_i \s}  + b_i e^{-i m_i \s}\ ,  \ee
where $a_i,b_i$ are complex coefficients.
The periodicity condition $z_i(\s + 2 \pi) = z_i(\s)$ 
implies that $m_i$ must be integer. 
To satisfy the constancy of $\L$ in \rf{ppl} we need to impose 
\be \la{cvb}
  \ss ( |a_i|^2 + |b_i|^2) =1 \ , \ \ \ \ \ \ \ \ 
  \ss ( m^2_i + w^2_i) ( a^*_i b_i e^{2i m_i \s} 
  +   a_i b^*_i e^{-2i m_i \s}) =0 
                 \ . \ee
In addition, we need to impose $ \ss |z_i|^2 =1$, i.e.
\be \la{yyy}
 \ss ( |a_i|^2 + |b_i|^2) =1 \ , \ \ \ \ \ \ \ \ 
  \ss ( a^*_i b_i e^{2i m_i \s} +   a_i b^*_i e^{-2i m_i \s}) =0   \ . \ee
It is easy to show that modulo the global  SU(3)  (subgroup of SO(6)) 
invariance 
of the system       \rf{L} 
or \rf{jk},\rf{ppl} the only non-trivial solution
of \rf{cvb},\rf{yyy}  is 
$b_i=0$ {\it or} $a_i=0$. In the former case 
($m_i$ may be positive or negative and $a_i$ can  be made 
real by $U(1)$ rotations) 
\be \la{iii}
z_i = a_i e^{i m_i \s} \ , \ \ \ \ \ \ \ 
\ \ \ \  \ss a^2_i=1 \ . \ee 
It may seem that one may  get a new solution 
if two of the windings $m_i$ are equal  while the third  is 
zero, i.e. (this is, in fact, the circular solution of \ci{ft2}) 
if \be 
z_1 = a \cos m \s \ , \ \ \ \ \ \ 
  z_2 = a \sin m \s \ ,\ \ \ \ \ \   \ z_3=\sqrt{1- a^2} \ , \ee
  but it can be transformed back into the form \rf{iii} by a global 
  SU(2) rotation.      
         
         \bigskip
             
 It is useful also to  rederive the solution \rf{iii}
 in slightly different way  
using  real coordinates $r_i,\a_i$, i.e.  starting with 
\rf{Laa},\rf{ank}.           
The potential $  w_i r^2_i + { v^2_i \ov r^2_i} $ 
in \rf{Laa} has a minimum, and that suggests 
that $r_i=$const may be a solution. 
That 
needs to be checked since   $r_i$ are constrained
to be on $S^2$.
The equations of motion that follow from \rf{Laa}
are 
\be
\label{Eqofm}
r_i''=-w_i^2 r_i + {v_i^2 \ov r^3_i}   -  \L  r_i \  , \ee
\be \la{lamm}
\L =  \sum_{j=1}^3
 \Big(r'^2_j-w_j^2r_j^2 + {v_j^2 \ov r_j^2} \Big) \, , \ \ \ \ \ 
 \ \ \ \ \ \ \ 
\sum_{j=1}^3 r^2_j =1 \ . 
\ee
They indeed have a solution  if 
\be \la{kop}
r_i(\s) =a_i =\const\ , \ \ \ \ \ \ \ 
w^2_i -  {v^2_i \ov a_i^4}  = \nn^2  =\const  \ , \ee
where $\n$ is an arbitrary constant 
(which may be positive or negative). Then it follows that 
 the Lagrange
multiplier in \rf{Laa} is thus constant on this 
solution 
\be \Lambda= - \nn^2 \ . \ee
As a result, we   obtain an interesting 3-spin {\it  generalization} 
of the   circular string solution  
 found  in \ci{ft2} (where  two out of 
 three spins where  equal).

Eq. \rf{kop} implies 
\be \la{pp}
a_i^2 =  {| v_i | \ov \sqrt{ w_i^2 - \nn^2 } }    \ ,
 \ \ \ \ \  \ \ \ \
 \a_i'= { v_i \ov a_i^2 }= { v_i \ov  | { v_i } | } 
\sqrt{ w_i^2 - \nn^2 } \equiv  m_i 
 \  ,   
 \ee
i.e.
\be 
\ \    \a_i = \a_{0i} + m_i \s     \ ,  \ee
where $m_i$ must be 
integers to satisfy the periodicity 
condition \rf{pep} and $\a_{0i}$ can be set to zero  by SO(2)
rotations.  Then 
\be\la{kol}
w_i^2 = m^2_i + \nn^2 \ , \ \ \ \ \ \ \  \ \ \ \ 
 v_i = a^2_i m_i  \ . 
\ee 
The constraints \rf{cv},\rf{cvv} or \rf{cve},\rf{cvve} give 
\be \la{kk} 
\k^2 =\ss  a^2_i (w_i^2 + m^2_i) =  2 \ss  a^2_i w_i^2 - \nn^2
  \ , \ \ \ \ 
\sum_{i=1}^3  a^2_i=1 \ , \ \ \ \ \ \ \ \ \ 
\sum_{i=1}^3  a^2_i  w_i m_i =0 \ , 
 \ee 
 or, equivalently, in terms of the energy and spins 
 (cf. \rf{enn},\rf{spins},\rf{genn})
 \be \la{onk}
 \E^2 = 2 \ss  w_i \J_i  - \nn^2 \ , \ \ \ \ \ {\rm i.e.} 
 \ \ \ \ \ 
 \E^2 = 2 \ss  \sqrt{m^2_i + \nn^2}\   \J_i  - \nn^2 \ , \ee
 \be\la{kpp}
 \ss {\J_i \ov w_i} =1 \  ,  \ \ \ \ \ {\rm i.e.}\  \ \ \ \ 
 \ss {\J_i \ov  \sqrt{m^2_i + \nn^2} } =1 \  ,\ee
 \be\la{uiy}
 \ss m_i \J_i =0 \ .  \ee
 We shall assume 
for definiteness that all $w_i$ and thus all $\J_i$ 
are non-negative. Then \rf{uiy} implies that  one 
of the three $m_i$'s 
 must have opposite sign to the  other two.

One can check directly that the only non-vanishing components of
the SO(6)  angular momentum tensor
$J_{MN} = \sql \int^{2\pi}_0 {d\s \ov 2 \pi} (X_M \dot X_N- X_N
\dot X_M)$ 
on this solution are indeed the Cartan ones 
$J_1=J_{12},\ J_2=J_{34},\ J_3=J_{56}$.

Since our aim is to express $\E$ in terms of $\J_i$ and $m_i$ as in \rf{ennp} 
the strategy is then to first  solve the condition 
\rf{kpp} in terms of $\nn^2$, determining it as a function of $\J_i$ and
$m_i$  and then substitute the result into \rf{onk}.
The condition \rf{uiy} may then be imposed at the very end. 

\bigskip

Let us first consider the special case of 
 $\nn^2=0$ (or $\Lambda=0$)   which 
 corresponds to a flat-space solution 
which can be  embedded into  $S^5$ by 
choosing the free radial parameters of a 
 circular string to satisfy the condition 
$\ss a^2_i =1$.  As follows from \rf{kol} for
 $\nn^2=0$ we  find  that all frequencies must be integer $w_i= |m_i|$, 
 e.g., 
\be \la{hh}
w_1 = -m_1 > 0  \ , \ \  
\ \ \ \   w_2 = m_2> 0  \ , \ \ \ \ \ \ w_3= m_3 > 0  \ , 
\ee
so that 
 the solution is a combination of the 
left and right moving waves
(here we use complex combinations of coordinates in 
\rf{emb})\foot{This may look as
 an example of a 
``flat'' or ``chiral''   
 solution of  $O(N)$ sigma model 
 that trivially 
 satisfies  the equations of motion following from \rf{SL}
 $\del_+ \del_-  X^M + \del_+ X_N \del_- X_N \ X^M =0$
since  $\del_+ X^M=0$ or $\del_- X^M=0$.
But one still needs to impose  the Virasoro constraints, 
and that implies that we need a particular 
combination of left and right
moving modes.}
\bea  
\label{hhh}
\XX_1 =a_1 e^{ i m_1 (\s - \tau)} \ , \ \ \ \
 \XX_2 =a_2 e^{ i m_2 (\s + \tau)} \  , \ \ \ 
 \XX_3 =a_3 e^{ i m_3 (\s + \tau)} \  , \ \ \ \  \ss a^2_i=1 \ . 
\eea
In the case of \rf{hh} we get from \rf{onk}--\rf{uiy}
\be 
 \E^2 = 2 \ss  |m_i|  \J_i   \ , \  \ \ \ \ \ \ 
  \ss {\J_i \ov |m_i|}
 =1 \ , \ \ \ \ \ \ \ \ 
 \ss m_i \J_i =0 \  .
   \ee
This  corresponds to a very special point 
in the moduli space of solutions. 
For fixed $m_i$, we get two  constraints on $\J_i$, and the energy is
given by the standard flat-space linear Regge relation.
For the choice \rf{hh} we end up with 
$|m_1 | \J_1 = m_2 \J_2 + m_3 \J_3$ (where $\J_2$ and $\J_3$ are
related via $\ss {\J_i \ov |m_i|}
 =1$) 
  and thus 
$\E^2 = 4  |m_1|  \J_1$. 
 Clearly, the energy of 
  this ``flat'' solution  does not have 
 a regular expansion  in $1\ov \J^2$ (cf. \ci{ft2}) and thus 
it  cannot be directly  compared to  some anomalous 
dimension on the SYM side.

 \subsection{
 Energy as function of the spins}
 
 Now let us turn to the genuinely ``curved''  ($\nn\not=0$) 
 solutions which will have indeed  a  regular  expansion 
 of  the energy  for  large spins as was in
  the case of the circular
 solution of \ci{ft2}.
  
 In the   3-spin case one is first to solve the equation 
 \rf{kpp} to determine $\nn$.  The solution of this 
 algebraic equation  cannot be written down 
  explicitly for generic $\J_i$ but one can find it 
  as a power series in large $\J=\ss \J_i$ expansion
  as in \ci{ft2,afrt} ($\J \gg 1$) 
\be 
\nn^2=\J^2  - \ss m^2_i  {\J_i\over \J} 
+...\ , \ \ \ \ \ \ \ \ 
 \E^2= \J^2  +  \ss m^2_i  {\J_i\over \J} 
+...\  , 
\ee 
where  $\J \equiv \J_1+ \J_2 +  \J_3 \gg 1$, 
 and  thus 
\be \la{epon}
\E = \J  + {1\over 2\J } \ss m^2_i  {\J_i\over \J} 
+...\ .
\ee
As in the previous examples in \ci{ft2,ft4,afrt}, 
here  the energy thus  admits a regular expansion in
 $1\ov \J^2$=${\lambda
\ov J^2}$
\be \la{epo}
E = J  + {\l\over 2J } \ss m^2_i  {J_i\over J} 
+...\ ,  \ \ \ \ \ \ \ \ \ \ \ \ss m_i J_i=0\ . 
\ee
Hence  it should be possible
 to match, as in \ci{mz2,afrt,bfst}, the coefficient 
 of the ${\cal O}(\l)$ term in \rf{epo}
 with the 1-loop 
 anomalous dimensions of the corresponding SYM operators
  determined by a
 special 3-spin case  of the integrable SU(2,2$|$4) spin chain 
 of \ci{bes2}.
  The simplicity of the expression \rf{epo}
 suggests that one may  be able to establish the 
 correspondence with  particular solutions of the Bethe ansatz
 equations in a relatively direct  way, as was the case in 
 \ci{mz2}
 for the $J_1=J_2,\ J_3=0$  and in \ci{mz3}
 for the $J_1=J_2,\ J_3\not=0$  
  circular solutions of \ci{ft2}.
 
 Let us now look at some special cases. 
If $\J_2=\J_3=0, \ a_2=a_3=0$, i.e. in the one-spin case,  
 we have a solution  if $w^2_1= \nn^2$, i.e. $m_1=0$
and $J_1= w_1$, and then $\E= \J_1$. 
This  is simply the point-like geodesic case:
    for $m_1=0$ there is no $\s$-dependence in $X_i$. 

In the two-spin case $\J_3=0, \ a_3=0$
the equation  \rf{kpp} for $\nn^2 $ becomes a quartic equation
\be \la{nea}
{\J_1 \ov  \sqrt{ m^2_1 + \nn^2 } } + 
{\J_2 \ov \sqrt{ m^2_2 + \nn^2 }} =1 \
.
\ee
Its simple explicit  solution is found in the case when 
$\J_1=\J_2$, i.e. $
  a_1= a_2 = {1\ov \sqrt {2}}$, $m_2=-m_1\equiv m  > 0$:
\be \la{uio}
\nn^2 =\J^2 -m^2 \ , \ \ \ \ \  \J\equiv \J_1 + \J_2 = 2 \J_1 \ , \ee
so that 
\be 
\E^2=\J^2+ m^2 \ . \ \ee
This is the same $\E(\J)$ relation 
 as for the 2-spin circular solution 
 of \ci{ft2}. In fact, as was already mentioned above, 
 the two solutions are {\it equivalent}:
 here we have 
 \be \la{hoh}
 \XX_1 = {1\ov \sqrt {2}} e^{ i w \tau - i m \s} \ , \ \ \ \ 
 \XX_2 = {1\ov \sqrt {2}} e^{ i w \tau  + i m \s} \ , \ee
 which is related to the solution in \ci{ft2} by an SO(4) rotation:
 \be\la{doo} 
 X'_1 = {1\ov \sqrt {2}}(X_1 + X_2)\  , \ \ \ \ \ \ \ \ 
 \ \ X'_2 ={1\ov \sqrt {2}}( -X_1 + X_2 ) \ . \ee
In the general case of two  unequal spins we can again  solve \rf{nea} 
in the limit of large 
$\J_1,\J_2$ (for fixed $m_1,m_2$), 
getting the special case of \rf{epo} with $m_1 \J_1+ m_2\J_2=0$, 
$\J_3=0$, i.e. 
\be \la{eppo}
\E = \J  + { m_2 ( m_2 + |m_1|) \J_2 \over 2\J^2 } 
+...\ = \J-\frac{m_1m_2}{2\J}+...\, .
\ee
In  another  special case  when two out 
of three non-vanishing spins are  equal, e.g., $ \J_2=\J_3$, and  
with  $m_1=0, \ m_2=-m_3=m$   we get from \rf{epo}
\be \la{ipo}
\E = \J  + { m^2 \J_2 \over \J^2 } 
+...\ =  \J +\frac{m^2}{2\J}+...\, .
\ee
This  is 
the same as the expression for the 
circular 3-spin solution ($\J_1\not=0,$ $\J_2=\J_3$) in \ci{ft2}.
  Indeed, for any values of 
 $J_1, J_2=J_3$
 the two solutions  are related by a global 
rotation in $X_2,X_3$ directions as  in \rf{doo}, 
converting $e^{i m_2 \s}$ into $\cos m_2 \s$ and $\sin m_2 \s$.

\bigskip

To summarize, 
we have shown that the constant-radius solutions of the NR system 
represent a simple generalization of the circular  2-spin and 3-spin
solutions of \ci{ft2}. This opens up a possibility 
of a direct comparison to SYM one-loop anomalous dimensions in the
(i) 2-spin sector with unequal spins (cf. \ci{mz2})
and 
(ii) general 3-spin sector (cf. \ci{mz3}). 

\bigskip

\subsection{Quadratic fluctuations and stability}

Let us now  study  small fluctuations near the 
solutions of section 3.1. This  will generalize 
(and simplify) the discussion in \ci{ft3} in the case of the special 
$J_1=J_2$ 3-spin solution  and will 
clarify  the conditions of stability of our new   solutions. 
One application of this analysis would be  to compute
 the 1-loop \sm\ correction to the classical energy \rf{epo} 
 and to show that it is indeed suppressed by extra power of $1/J$ 
 as in the special case considered in \ci{ft3}. 
 Another would be  to find the spectrum of excited string states 
 carrying the same charges as the ``ground-state'' classical solution
 as these may be possible to compare to the corresponding spectrum of anomalous
 dimensions  on the SYM side (as was done 
 for the special $J_1=J_2,\ J_3=0$ case in \ci{mz2}).

It is straightforward to find the quadratic fluctuation Lagrangian 
by  expanding near the solution \rf{iii} or \rf{kop}--\rf{kol}.
We shall follow the discussion in section 2 of 
\ci{ft3} where the special case of circular solution with two equal
 spins was considered. 
Using three complex combinations of coordinates in \rf{emb} and
expanding
$\XX_i \to \XX_i + \td \XX_i$ 
the sigma model action \rf{SL} near the classical solution 
\be 
\XX_i = a_i e^{i w_i \tau + i m_i \s} \ , \ \ \ \ \ \ \ 
w^2_i = m^2_i + \nn^2 \ , \ \ \ 
\ss a^2_i=1 \ , \ \ \  \ss a^2_i w_i m_i =0 \ , \ee
we  find the following Lagrangian for the quadratic fluctuations 
(see \ci{ft3}) 
\be \la{flu}
\td L = - \ha \del_a \td \XX_i \del^a \td \XX^*_i  + \ha \L 
\td \XX_i  \td \XX^*_i \ , \ee
where  $\L= -\nu^2$ and 
$\td \XX_i $ are subject to the constraint\foot{Note that 
imposition of
Virasoro  constraints on the fluctuations is not necessary 
in order to determine the non-trivial part of the fluctuation
spectrum \ci{ft3}.} 
\be \la{koi} 
\ss ( \XX_i \td \XX^*_i + \XX^*_i \td \XX_i ) = 0 \ . \ee 
To solve this  constraint we set 
\be \la{sett}
\td \XX_i = e^{i w_i \tau  + i m_i \s} Z_i (\tau,\s) \ , \ \ \ \ 
\ \ \  Z_i = g_i + i f_i \ , \ee
so that  \rf{koi} becomes 
\be
\la{oyy}  \ss a_i  g_i  = 0 \ . \ee
After integrating by parts, \rf{flu} takes the form 
(cf. \ci{ft3})
\be \la{fle}
\td L = 
 \ss \bigg[ \ha (\dot f^2_i   + \dot g^2_i - f'^2_i - g'^2_i )
-  2 w_i f_i \dot g_i  + 2 m_i f_i g'_i \bigg] \ . \ee
To solve \rf{oyy}  we may apply a global rotation to $g_i$, 
\ $\bar g_i = M_{ij} (a) g_j$,  that 
transforms $\ss a_i  g_i $ into $\bar g_1$  and set the latter 
 to zero in the resulting Lagrangian \rf{fle}. 
 Equivalently, we may solve \rf{oyy} for $g_1$ and substitute the
 result  into \rf{fle}.
 
For simplicity,  let us first consider the  2-spin case when (cf.
\rf{nea}) 
\be\la{oll}
a_1^2 + a^2_2 =1   \ , \  \ \  a_3=0\ ,
\ \ \ 
a^2_1 |m_1| w_1  - a^2_2 m_2 w_2  = 0 \ , \ \  \ \ \ 
w^2_1 - m^2_1 = w^2_2 - m^2_2 = \nn^2 \ . \ee 
We shall  assume that $m_1 < 0, \ m_2 > 0$.
These relations allow us to express $a_1$ and $a_2$
 in terms of 
$m_1,m_2$ and $\nu$
\be \la{kpl}
a_1^2 = { m_2 \sqrt{ m^2_2 + \nu^2} \ov
 |m_1| \sqrt{ m^2_1 + \nu^2} +m_2 \sqrt{ m^2_2 + \nu^2} } \ , \ \ \
 \ 
 \ \ \
 a_2^2 = { |m_1| \sqrt{ m^2_1 + \nu^2} \ov
 |m_1| \sqrt{ m^2_1 + \nu^2} +m_2 \sqrt{ m^2_2 + \nu^2} } \ . 
\ee
In this case the fluctuations in the  $i=3$ direction decouple, and
we 
 find the following Lagrangian for the remaining 3 fluctuations 
 $g_2,f_1,f_2$ (e.g. solving \rf{oyy} for $g_1$ 
 and rescaling $g_2$)
 $$ \td L =  \ha (\dot f^2_1   + \dot f^2_2 
 + \dot {g}^2_2 - f'^2_1  -f'^2_2 -    g'^2_2 )  $$
 \be\la{uuu}
  + \ 
 2 (  a_2 w_1 f_1  -  a_1 w_2 f_2 )\dot g_2 
 -  2 (  a_2  m_1 f_1  - a_1 m_2 f_2 ) g'_2  
 \ . \ee
 Solving the resulting equations of motion  for 
 $F_q=(f_1,f_2,g_2)$
 using the ansatz  (see \ci{ft3}) 
 $F_q =  \sum_{s,n}   A_{q,s,n} e^{i \omega_s \tau + i n \s}$ 
 we find the
 following 
  characteristic equation of the frequencies $\omega$
 \be \la{bbb}
 (\om^2 - n^2)^2  - 4  a^2_2 ( w_1 \om - m_1 n)^2 
 - 4 a^2_1  ( w_2 \om - m_2 n)^2 =0 \ .  \ee 
 This is a quartic equation for $\om$,
 and the stability condition is 
 that all four roots should be  real. 
 The solutions are obviously real for $n=0$ so instability 
 may appear only for $n=\pm 1, ...\ $. 
   In the special case of the 2-spin circular
 solution of \ci{ft2}, i.e. $w_1=w_2=w, \ -m_1=m_2=m, \ 
 a^2_1  = a^2_2   = \ha $ we get 
\be \la{soo}
 (\om^2 - n^2)^2  - 4 w^2  \om^2  - 4 m^2  n^2  =0 \ ,  \ee 
which implies instability  when $ n^2 - 4 m^2 < 0$, 
i.e. for the modes  with 
$n= \pm 1, ...,\pm (2m-1)$ \ci{ft3}. 
 
  For generic $a_1,a_2,m_1,m_2$ and small enough $n$ 
  one finds that two of the four roots are 
  complex (with non-zero real part).\foot{
  For example, 
  setting $m_1=-1, \ m_2=2 , \ n=1$  one 
  gets complex solutions for
  $\nu$ from 0 to $1000$.
  This implies instability of  ``asymmetric'' solutions
  with $|m_1|\not=m_2$.} 
 In spite of the  instability it is  useful to work out the
 spectrum of frequencies and the stability condition 
 in the limit of large spins (i.e. large  $\nu$)
   since the resulting energies   may be
    compared to SYM theory.
 First,  let us consider  the case of equal spins
 ($-m_1= m_2=m$).  Eq. \rf{soo} implies that  \ci{ft3} 
 \be 
 \om^2_{\pm} = n^2 + 2 \nu^2 +  2 m^2  \pm 2 \sqrt{ ( \nu^2 +
 m^2)^2 
 + n^2 ( \nu^2 + 2 m^2) } \ , \ee
 so that the large $\nu$ expansion gives 
 (for the lower-energy modes)
 \be \la{kkk}
 \om^{(\pm)}_{-} = \pm { 1\ov 2\nu} n \sqrt{n^2-4m^2} 
 + {\cal O}( { 1\ov \nu^3}) 
 \ . \ee 
 Then  the ``one-loop''
 contribution to the energy of rotating string 
  from (a pair of) such modes
 is (here $\k^2 = \nu^2 + 2 m^2$, $  J =J_1+ J_2 = \sql 
  \sqrt{ \nu^2 + m^2}$; 
    $m=k$ in the  notation of  \ci{ft3})
 \be \la{huh}
 \Delta E_n = { 1 \ov \kappa}  \ 2 |\om_{-}|
 = { 1\ov  \nu^2} n \sqrt{n^2-4m^2}  + {\cal O}( { 1\ov \nu^4}) 
= { \l\ov  J^2} n \sqrt{n^2-4m^2}  +{\cal O}( { \l^2 \ov J^4})
\ . \ee
This expression was indeed reproduced in \ci{mz2} 
(for $m=1$) as the 1-loop  anomalous dimension of excited states 
on the SYM side  (corresponding to a particular Bethe
root distribution for the Heisenberg  spin chain).

In the general $(m_1, m_2)$ case,  expanding \rf{bbb} at 
large $\nu$ assuming\foot{There are also two other 
frequences  for which $\om^2
\to 4
\n^2$ at large $\nu$.} 
$\om = {\cal O}( {1\ov \nu})$ we find the following generalization of 
\rf{kkk}
\be \la{kuk}
 \om_{-} =  { 1\ov 2\nu}
n \bigg[ 2a_2^2 m_1   +  2  a_1^2 m_2  
\pm   \sqrt{n^2-  4   a^2_1 a^2_2 (m_1 -m_2)^2  } \bigg]
 + {\cal O}( { 1\ov \nu^3}) 
 \ ,  \ee 
 where $a^2_1+ a^2_2=1$. 
 This reduces to \rf{kkk} in the equal-spin case when 
 $a^2_1=a^2_2=\ha, \ m_1=-m_2$. 
 Stability condition is
  $n^2\geq   4   a^2_1 (1-a^2_1)  (m_1 -m_2)^2 $.
 If we recall that we have the constraint $m_1 J_1 + m_2 J_2=0$ 
 where $J_i = a^2_i \sqrt{m^2_i + \nu^2}$ 
 one may wish to solve it in the large $\nu$ limit getting 
 $a^2_1 m_1 + (1-a^2_1) m_2=0$, i.e. $a^2_1= { m_2 \ov m_2-m_1}$, 
 $1-a^2_1= -{ m_1 \ov m_2-m_1}$, giving the condition 
 $n^2 \geq 4|m_1m_2|$, which implies the existence of 
 unstable  modes  with $n^2  < 4|m_1m_2|$. 
 
One should be able to reproduce the analog of \rf{huh} in the case
of \rf{kuk}, i.e. (here we assume $|m_1| >  m_2$)
\be \la{heh}
 \Delta E_n = { \l\ov  J^2} n  
\bigg| 2(|m_1| - m_2)  -   \sqrt{n^2-4|m_1| m_2 } \bigg|
+{\cal O}( { \l^2 \ov J^4})
\ . \ee
 on the gauge theory side. 
 

\bigskip 

It is  straightforward to extend 
 the above discussion 
to the 3-spin case, i.e. when $a_3$ is non-zero. 
This will give a generalization of the spectrum found in the 
$(J_1, J_2=J_3)$ case in \ci{ft3}; as in that special case, 
 there should then be 
a range of parameters for which the solution is stable. 
The generalization of the eq.\rf{bbb} to the 3-spin case is\foot{
It can be found, e.g.,  by adding the constraint \rf{oyy} 
to the Lagrangian \rf{fle}  and   solving the corresponding equations
of motion.}
$$
(\om^2 - n^2)^4 - (\om^2 - n^2)^2 \big[(a^2_2+ a^2_3) \Om^2_1 
+ (a^2_2+ a^2_3) \Om^2_2 +(a^2_1+ a^2_2) \Om^2_3 \big]
$$
\be \la{aaa} 
+ \ a^2_3 \Om^2_1 \Om^2_2 + a^2_2 \Om^2_1 \Om^2_3 + 
a^2_1 \Om^2_2 \Om^2_3 =0 \ , \ee
where 
\be
 \Om_i \equiv 2 (w_i \om - m_i n) \ , \ \ \ \ \ \ \ \  
 w_i =\sqrt{ m^2_i + \nu^2} \ .  \ee
Setting $\Om_3=0,\ a_3=0$ we indeed go back to \rf{bbb}.
This equation gives  8 characteristic frequencies. 
Solving the equations for $a_2,a_3$ in terms of $a_1$
and $w_i =\sqrt{m^2_i + \nu^2}$ we get the following 
generalization of 
\rf{kpl}  
 \be \la{eii}
a_2^2 = - { m_3 w_3 \  (1- a^2_1) + 
m_1  w_1   \ a^2_1  \ov
 m_2 w_2 - m_3 w_3  } \ , \ \ \ \ \ \ 
 a_3^2 = {  m_2 w_2 ( 1 - a^2_1)  + m_1 w_1   a^2_1  \ov
  m_2 w_2 - m_3 w_3  } \ .  
\ee
Concentrating then on those  frequencies that scale as 
\be \om ={\vom\ov \nu} + {\cal O}({1\ov \nu^2})\ , \ \ \ \ \  \nu \gg 1    \ee
we get the following equation for the leading part
 of eq.\rf{aaa}
\be\la{kkyk}
  A + B a^2_1=0 \ , \ee 
$$
A= \big[4 (\vom - n m_3)^2 - n^4 \big] 
\bigg[ 4 [\vom - n (m_2 + m_3)]^2 
- n^2(n^2 + 4 m_2  m_3) \bigg] \ , 
$$ 
\bea
\nonumber
B&=& 4    (m_1 - m_2) (m_1- m_3) n^2 \\
\nonumber
&\times&
\bigg[ 12 \vom^2  - 8 n (m_1 + m_2 + m_3) \vom  
+ 4 n^2 ( m_1 m_2 +  m_1m_3 + m_2 m_3)  + n^4 \bigg] \ . 
\eea
Stable solutions arise in the range of the parameters $m_1,m_2,m_3$
such that \rf{kkyk}  has real roots $\bar \omega $ for all 
integer $n$.
The general stability condition on $m_1,m_2,m_3$ and $a^2_1$ 
appears to be  complicated, but one can
find particular values of $m_1,m_2,m_3$ for which 
 the solution is
stable.

For example, setting $m_1=0,\ m_2=-m_3=m$,  so that  
$a_1\equiv a,  \ a^2_3=a^2_2= \ha ( 1 - a^2) $, 
  which  is the case of the 
 3-spin solution of \ci{ft2}, 
  $\J_1 = a^2 \nu, \ \J_2=\J_3= \ha ( 1 -
a^2)\sqrt{m^2 + \nu^2} $, 
  we find, in agreement with \ci{ft3}\foot{In the notation of 
  \ci{ft3}   $a=\cos\g_0, \ m={\rm k}$.}
 \be 
 \vom^2 = { 1 \ov 4} n^2 m^2 \bigg[  {n^2\ov m^2}   - 2  + 6  a^2 
 \pm 2  \sqrt{ (3 a^2-1)^2  +  
 4a^2 ({n^2\ov m^2} -1 ) }\bigg] \ . 
 \ee 
 The condition of stability, i.e. $\vom^2 \geq 0$ 
 is obtained  by demanding that 
 $(q^2 - 4 ) ( q^2 - 4 a^2 ) \geq 0$ and 
 $  (3 a^2-1)^2  +  4a^2 (q^2- 1)    \geq 0$, where
 $q\equiv { n\ov m}$. The stability condition is satisfied 
 if $q^2 \geq 1$  and $a^2 \geq { 1 \ov 4}$, 
 which applies to all modes if $m=1$ as in 
 \ci{ft3}.  For $m = 2$ the potentially unstable  mode
 is $n=\pm 1$   having $q^2= { 1 \ov 4}$. Then to have stability 
 we need to demand $a^2 \geq { 1 \ov 16}$ as well as 
 $ { 1 \ov 16}  \leq  a^2 \leq { 1 \ov 6} ( 3 - \sqrt 5)$ or 
 $  { 1 \ov 6} ( 3 + \sqrt 5) a^2  < 1 $. 
 Similar conditions on $a$ are found for higher values of
 $m$.
 
If  instead we set $m_1=m_2$ (or $m_1=m_3$) in \rf{kkyk} 
 we find 
\be
\vom= n (  m_3 \pm \ha n) \ , \ \ \ \ \ \ 
\vom=  n \big[ (m_2 + m_3 ) \pm \ha \sqrt{ n^2 - 4 |m_2m_3 | } \big] 
\ee
implying that  modes with $n^2 <  4 |m_2m_3 | $ are unstable 
irrespective of the value of  $a_1$, just like in the 
2-spin case \rf{kuk}.

 \setcounter{equation}{0}

\section{More general ``non-constant''  solutions\\
  of the Neumann-Rosochatius  system}

\subsection{NR equations  in ellipsoidal coordinates}

Analogously to the case of the Neumann system in \ci{afrt}
we can rewrite the 
equations of motion following from \rf{Laa}
in the ellipsoidal coordinates $(\zeta_1,\zeta_2)$ 
which are 
introduced as 
\bea
r_i=\sqrt\frac{(w_i^2-\zeta_1)(w_i^2-\zeta_2)}{\prod_{j\neq i} w_{ij}^2}\, ,
~~~~~~~~~~~
w_{ij}^2 = w_i^2 - w_j^2 \ . 
\eea
Expressing the integrals of motion (\ref{nti}) in terms of 
$\zeta_a$ one finds the following separable 
system of  equations
\bea
\label{emel}
\left(\frac{d\zi}{d\s}\right)^2=-4\frac{P(\zi)}{(\zi-\zii)^2} \, , ~~~~~~~
\left(\frac{d\zii}{d\s}\right)^2=-4\frac{P(\zii)}{(\zi-\zii)^2}\, ,
\eea
where $P(\zeta)$ is 
\bea\nonumber
P(\zeta)&=&(\zeta-b_1)(\zeta-b_2)(\zeta-w_1^2)(\zeta-w_2^2)(\zeta-w_3^2)\\
\nonumber
&+&v_1^2 (\zeta-w_2^2)^2(\zeta-w_3^2)^2+v_2^2
(\zeta-w_1^2)^2(\zeta-w_3^2)^2
+v_3^2 (\zeta-w_1^2)^2(\zeta-w_2^2)^2\ . 
\eea
Here $b_{1,2}$ are the constants of motion which can be expressed
in terms of  the original integrals $I_i$ in \rf{nti}. 
The Hamiltonian
of the NR system reduces then to 
\bea
H=\frac{1}{2}\Big[\sum_{i=1}^3(w_i^2+v_i^2)-b_1^2-b_2^2\Big] \, .
\eea
As in the  Neumann case, $P(\zeta)$ is the fifth order 
polynomial which defines a  hyperelliptic curve
$s^2+P(\zeta)=0$. However, with non-zero $v_i$ 
the positions of the roots
get shifted. The general solution of the eqs. (\ref{emel})
can be given in terms of theta-functions associated to the
Jacobian of the hyperelliptic curve.  

We will not consider the problem of solving the equations 
(\ref{emel}) in full generality, rather we will 
treat the simplest case 
of the vanishing integral $v_3$. As one can see, 
 for 
$v_3=0$ the value  $\zeta=w_3^2$ is a root of $P(\zeta)$
and then the 
 NR system can be solved in terms of elliptic functions.

\subsection{Two-spin solution of the  NR system}

If $v_3=0$ we may set $\a_3=0$ and further assume
that  $ r_3=0$ (see \rf{ank}) 
which brings us to the two-spin case.
In terms of the ellipsoidal coordinates the two-spin solution 
arises in the limit $b_2\to w_3^2$.\footnote{To be specific we will treat the case 
of the folded string (cf. \ci{afrt}), analysis of the circular string solution
 is very similar.} It is convenient to perform
the following change of variables
$\zeta_a \to \xi_a$  (see \cite{afrt} for details)
\bea
\zeta_1\to w_2^2-(w_2^2-b_1)\xi_1 \, , 
~~~
\zeta_2\to w_3^2-(w_3^2-b_2)\xi_2\ . 
\eea
Then we find  that the first equation in 
 (\ref{emel}) reduces to 
\bea
\label{curve}
(\xi')^2=4w_{21}^2\xi(1-\xi)(1-\st\xi)-4v_1^2\xi^2-4v_2^2(\frac{1}{\st}-\xi)^2
\, , ~~~~~~\xi\equiv \xi_1\, .
\eea
Here $\st\equiv (w_2^2-b_1)/w_{21}^2$ is the modulus
of the elliptic curve.  
The variable $v_2$ can be eliminated using 
$v_2^2=v_1^2w_1^2/w_2^2$. Thus we get a 
one-parameter  family of solutions (parametrized by 
the additional parameter $v_1$). 

It is possible to reduce the elliptic curve 
corresponding to (\ref{curve}) to the standard Jacobian form,
but the new modulus 
$k$ appears to be a  rather complicated function of $\st,w_1,w_2,v_1$. 
Indeed, we get 
\bea
(\xi')^2=4w_{21}^2\st(\xi-e_0)(\xi-e_1)(\xi-e_2)\ , 
\label{cub}
\eea
%
where 
\be v_1^2= w_{ 21}^2 \st^3\ e_0e_1e_2\ ,\ \ \ 
\ \ \ \ 
v_1^2-v_2^2+\kappa^2-w_2^2= -w_{21}^2 \st^2(e_0e_1+e_0e_2+e_1e_2)\, .
\ee
After  the change of variable $\xi= e_{10}\eta^2+e_0$,
where $e_{nm}\equiv e_n-e_m$, 
eq. (\ref{cub}) becomes
\bea
(\eta')^2=w_{21}^2\st  e_{20} (1-\eta^2)(1-k\eta^2)\, ,
~~~~~~~k=\frac{e_{10}}{e_{20}} \, .
\eea
Thus, a solution obeying the condition $\eta(0)=0$ reads as 
\bea
\label{soleta}
\eta(\s) =\mbox{sn}\Big( \sigma \sqrt{w_{21}^2\st e_{20}}, k\Big)\, .
\eea
The radii of the embedding coordinates in \rf{hah} then 
are 
\bea
r_1^2(\s) =1-\st(e_0 + e_{10}\eta^2)\, , \ \ \ \ \ 
~~~~~r_2^2(\s) 
=\st(e_0+e_{10}\eta^2)\, .
\eea
Note that this is  the most general two-spin solution of the NR system.
In the present case, we require in addition that $\eta$ should 
 be periodic,
$\eta(\s+2\pi )=\eta(\s )$, which gives 
\be
\frac{\pi}{2}\sqrt{w_{21} ^2\st e_{20}}=\ellK(k)\, ,
\label{peri}
\ee
where $\ellK(k)$ (and $\ellE$ and $\Pi$ appearing 
 below) are the  standard  elliptic functions defined, e.g., 
  in \ci{afrt}.

Since for  the  periodic solutions 
 $ \varphi_{1,2}(\s+2\pi )= 
\varphi_{1,2}(\s )+2\pi m_{1,2}$ we have also 
the condition \rf{okl}, 
 we can trade the parameters $v_1,v_2$ 
 for the  two integers $m_1,m_2$. 
Using the explicit solution 
(\ref{soleta}) one can compute the integrals in eq. (\ref{okl}) with 
the result 
\bea
m_1=\frac{v_1 }{(1-\st e_0)\ellK(k)}\Pi\Big(\frac{\st e_{10}}
{1-\st e_0}, k\Big) \, ,
\ \ \ \ \ \ \ \ \ \ 
m_2= \frac{v_2 }{\st e_0\ellK(k)}\Pi\Big(\frac{e_{01}}{e_0}, k\Big) 
\, . \la{kkj}\eea
For given non-zero integers $m_i$ these are highly transcendental
equations on $v_1,v_2$. 
Computing the spins we get
\bea
{\cal J}_1&=&w_1 - w_1 e_0 \Big(1 + k - \frac{\ellE(k)}{\ellK(k)} \Big)\ , \\
{\cal J}_2&=&w_2e_0\Big(1+k-\frac{\ellE(k)}{\ellK(k)} \Big) \ . 
\eea
Finally,  the energy is given by 
\bea\la{pj}
{\cal E}^2=\kappa^2=w_1^2+\st w_{21}^2-v_1^2\Big(1+\frac{w_1^2}{w_2^2}\Big)\, .
\eea
Note that due to the extra condition \rf{cvve}, i.e. 
$v_1w_1=-v_2w_2$, 
 the solution exists only if $J_1$ and $J_2$
are related in a certain way.

The above 
system of equations \rf{peri}--\rf{pj}, determines
the energy $E$ parametrically 
 as a function of the R-charges $J_1=\sql \J_1,\
J_2=\sql \J_2$ and winding numbers $m_1,m_2$. 
This system  is rather complicated to allow
for an explicit formula for 
$E=\sql \E({J_1\ov \sql} ,{J_2\ov \sql}; m_1,m_2)$.
Nevertheless, 
we hope that it might
  be possible to directly match this system 
(its leading ${\cal O}(\l)$ or the ``one-loop'' approximation) 
onto the corresponding equations
governing the algebraic Bethe ansatz 
(for a particular choice of the
Bethe root distribution)
for the anomalous dimensions of the corresponding operators 
on the gauge theory side,
as was done in the $v_i=0$ case in \ci{mz2,bfst}. 



\setcounter{equation}{0}
\section{Rotating strings in  $AdS_5\times S^5$ 
}

Let us now generalize the discussion of sections 2 and 3
to the case when the string can rotate in both $AdS_5$ 
and $S^5$. 
For that we need to supplement the 
$S^5$ rotating string ansatz \rf{emb}  by the 
similar $AdS_5$ one 
$$
\Y_0\equiv Y_5+i Y_0= \zz_0(\s ) e^{i \w_0\tau }\ , \ 
 $$
\be \la{adr}
\Y_1\equiv Y_1+i Y_2= \zz_1(\s ) e^{i \w_1\tau }\ ,
\ \ \ \ \  \ \ \Y_2\equiv  Y_3+i Y_4= \zz_2(\s ) e^{i \w_2\tau } \
, 
\ee 
where now (generalizing the ansatz  considered in 
\ci{ft2,afrt}) 
 $\zz_r=(\zz_0,\zz_1,\zz_2)$ are complex, and  because of
the condition
$\eta_{MN}Y^M Y^N=-1$, their real  radial parts 
 lie on a hyperboloid ($\eta_{rs}=(-1,1,1)$) 
\be\la{dos} 
\zz_r = \rr_r e^{i \b_r} \ , \ \ \ \ \ \  \ \ \ \ \ 
\eta^{rs}\rr_r\rr_s \equiv -\rr_0^2+ \rr_1^2 + \rr^2_2  
=-1\ . 
\ee
In the previous sections  we had $\rr_0=1, \ \rr_1=\rr_2=0, \ \b_r=0.$
To satisfy the  closed string periodicity conditions  we need, as
in
\rf{pep},  
\be \la{tqr}
\rr_r (\s + 2 \pi)= \rr_r (\s) \ , \ \ \ \ \ \ 
\b_r (\s + 2 \pi)= \b_r (\s)  + 2 \pi k_r \ , \ee
where $k_r$ are integers. Comparing \rf{adr}  to \rf{relx} we
conclude that 
the $AdS_5$ time $t$ and the angular coordinates $\p_1,\p_2$
are related to $\b_r$ by 
\be \la{tte}
t = \w_0  \tau  + \b_0 (\s) \ , \ \ \ \ \ 
\p_1 = \w_1  \tau  + \b_1 (\s) \ , \ \ \ \ \
\p_2 = \w_2 \tau  + \b_2 (\s) \ . \ee 
We shall require  the time coordinate $t$
to  be single-valued (we are considering a universal cover of
$AdS_5$), i.e. ignore windings in time direction  
and will also rename $\w_0$ into $\k$, i.e. 
\be  k_0=0 \ , \ \ \ \  \ \ \ \ \ \ \ \w_0 \equiv \k   \ . \ee 
The three O(2,4) Cartan generators (spins)  here 
are ($S_0=E, \ \w_r=(\w_0,\w_1,\w_2)$)
\be \la{cha}
S_r=\sqrt{\lambda }\ \w_r \int_0^{2\pi } {d\sigma\over 2\pi }\ 
\rr_r^2(\sigma )\equiv \sqrt{\lambda }\ \S_r \ . 
\ee
In view  of (\ref{dos}), they satisfy the relation
\be\la{hjh}
\sum_{s,r} \eta^{sr} {\S_r\ov \w_s} =-1 \ , \ \ \ \ \ \ \ {\rm i.e.}
\ \ \ \  \ \ \ 
{\E\over \kappa } - {\S_1 \over \w_1}  - {\S_2 \over \w_2}=1 \ . 
\ee
Substituting the above rotational ansatz into the $AdS_5$
Lagrangian 
(and changing overall sign) 
we find the analog of the 1-d  Lagrangian \rf{L} in the $S^5$ case
\be\la{tl}
\td \rL = \ha \eta^{rs} (\zz_r'{\zz_s^*}'-\w_r^2  \zz_r\zz_s^* )-
\ha \tilde \Lambda (\eta^{rs} \zz_r\zz_s^*   + 1) \ . 
\ee
Like its $S^5$ counterpart \rf{L}, this 1-d 
 Lagrangian is a special case of an $n=6$ Neumann system
now with  signature $(-++++-)$, and thus represents again 
an integrable system  (being related, as in \ci{afrt}, 
  to  a special 
euclidean-signature Neumann  model by an analytic continuation).  
 The  reduction of the total \adss\ Lagrangian on the rotation
 ansatz 
 is then given by the sum of \rf{L}  and \rf{tl}. 
Writing \rf{tl} in terms of $\rr_r$ and $\b_r$ 
 we find as in \rf{ank}
\be\la{ankk}
\b'_r  = { u_r\ov \rr^2_r}\ , \ \ \ \ \ \  \ \ \ u_r=\const \ , \ee
 so that finally we end up with 
\be\label{lagr}
\tilde \rL= \ha \eta^{rs} ({\rr_r'}{\rr_s'}
- \w_r^2 \rr_s\rr_s - {u_r u_s\over
\rr_r \rr_s}  )  - 
  \ha \tilde \Lambda (\eta^{rs} \rr_r\rr_s    + 1) \ , 
\ee
where, as above, we assume summation over $r,s$. 
Comparing this  to  the NR Lagrangian (\ref{Laa}),
we conclude  that  (\ref{lagr}) describes a system 
which is similar to
the Neumann-Rosochatius integrable system, but with an 
indefinite signature, i.e. $\delta_{ij}$ replaced by $\eta_{rs}$.

While  the  equations for $r_i$  and $\rr_r$ following,
respectively,  from 
\rf{Laa} and \rf{lagr} are decoupled, the variables of the two NR 
systems are mixed in the conformal gauge constraints 
\rf{cv},\rf{cvv} which now take the form (generalizing 
\rf{cve},\rf{cvve} where we had $\rr_0=1, \ u_r=0, \ \rr_a=0$)
\be \la{cvi}
 \rr'^2_0   + \k^2 \rr^2_0    + {u_0^2 \ov \rr^2_0} 
 = \sum_{a=1}^2 ( \rr'^2_a   + \w_a^2 \rr^2_a  +  
 { u_a^2 \ov \rr^2_a} )   +    \sum_{i=1}^3( r'^2_i   + w_i^2 r^2_i
  +  
 { v_i^2 \ov r^2_i} )    \ , 
\ee
\be \la{cvvi}
\ \k u_0   = \sum_{a=1}^2 \w_a  u_a   +  \sum_{i=1}^3 w_i v_i    \
,
\ee
where 
$  \rr_0^2 - \sum_{a=1}^2 \rr_a^2 =1  ,$ and  $
\sum_{i=1}^3 r^2_i =1 .$ 
We should also require the periodicity condition analogous to 
\rf{okl}
\be\la{ukl}
u_r \int^{2\pi}_0 { d\s \ov \rr^2_r(\s)} = 2 \pi k_r  \ .
\ee
Then $k_0$ implies that  we should set $u_0=0$
as a consequence of single-valuedness of the $AdS_5$ time $t$.

One  can then repeat the discussion of sections 2,3  and 
 in the present case, classifying   general solutions 
of the resulting NR system.
The resulting solutions generalize those of sect.4.2 in \ci{ft2} 
 where the integrals $v_i$ and $u_\rr$ 
were zero.

\subsection{Simple circular strings in $AdS_5$ }

Let us first  assume that the string 
is not rotating in  $S^5$ (i.e. $w_i,v_i=0, \ r_i=\const$) 
and consider the $AdS_5$ analog of the simplest circular solution 
of section 3 by demanding that $\td \L=\const$. 
The discussion is then exactly the same as
(a special case of that)  in section 3 with 
few  signs reversed. 
As in section 3.1, finding solutions with $\td \L=\const$ 
 turns out to be equivalent 
to looking for constant radii ($\rr_r=\const$)  configurations.
Then (cf. \rf{kop},\rf{kol}) 
\be \la{hop}
\rr_r=\const \ , \ \ \ \ \  \beta_{a} =k_{a}\s \ , \ \ 
\ \ k_0=0\ , \ \ \ \  u_0=0 \ , \ \ \ \
u_a = \rr_a k_a \ , \ee
 \be \ \w^2_0\equiv \k^2=  \td \Lambda \ , \ \ \ \ \ \ \ \ \ \ \ 
\w^2_a = k^2_a  + \k^2  \ , \ \ \ \ \ \ a=1,2 \ .  
\ee
The energy  as a function of spins is then obtained by solving 
the system of the  two equations that follow from the definition
of the charges \rf{cha}
 and  the  constraints \rf{cvi},\rf{cvvi}
 with $\kappa$ as a parameter (cf. \rf{kpp}--\rf{uiy})
\be
{\E\over\kappa } - {\S_1 \over \sqrt{ k^2_1 + \k^2}}  - 
{\S_2\over  \sqrt{k^2_2 + \k^2} } = 1\ , 
\label{cuat}
\ee
\be\la{nnb}
\k \E - \ha \k^2  = \sqrt{ k^2_1 + \k^2 }\ 
 \S_1  +  \sqrt{k^2_2 + \k^2}\  \S_2   
\ , \ \ \ \ \ \  k_1 \S_1  +  k_2 \S_2=0\ . 
\ee 
This implies 
\be
{ k^2_1\S_1 \over \sqrt{k^2_1 + \k^2}}  +  
{ k^2_2\S_2 \over  \sqrt{k^2_2 + \k^2} } = \ha \k^2 \ . 
\ee 
Considering the limit of large spins  $\S_a  \gg 1$, 
with  $k_a$ being fixed,  
we conclude that
 $\k = ( 2 k_1^2 \S_1 + 2 k_1^2 \S_1)^{1/3} + ...$ 
 and then  
\be
\E= \S_1 + \S_2 + {3\over 4} (2k_1^2\S_1+2k_2^2\S_2)^{1/3}+...
\ ,   \ee
or,  in view of  $k_1 \S_1 = -k_2 \S_2$
(treating $\S_1,\S_2$ and $k_1$ as independent data) 
\be\la{pqw}
\E= \S + {3\over 4} \big( 2k^2_1 \S  { \S_1\ov \S_2} \big)^{1/3}
+...\ , \ \ \ \ \ \ \ \  \S\equiv \S_1 + \S_2 \ .   
\ee
Using \rf{cha} this can be rewritten also as 
\be\la{qqw}
E= S + {3\over 4} (\l S)^{1/3}  \big( 2k_1^2 {S_1\over
S_2}\big)^{1/3}
+ ...\ .    
\ee
The case of  $k_1=-k_2=k$ when the two spins are equal 
$\S_1=\S_2=\ha \S$  is that of the the circular 
solution found in  \ci{ft2}  for which we get 
\be 
\E=  \S + {3\over 4}(2k\S)^{1/3} + ...\ . 
\ee 
As was shown in \ci{ft2}, this $k_1=-k_2$ solution is 
stable only for  small enough $\S$. 

The ``non-perturbative'' scaling of the subleading 
term  in \rf{qqw} with $\l$ precludes  one  from 
entertaining a possibility  of a direct comparison to
anomalous dimensions of the corresponding operators \ci{ft2}, i.e. 
$\bar \Phi D^{S_1}_{1+i2} D^{S_2}_{3+i4}\Phi $,  in  SYM theory, 
in contrast to what was found in the $S^5$ case. 

Let us now  see how this conclusion changes when we consider 
``hybrid'' solutions where the circular string rotates both in 
$AdS_5$ and $S_5$.

\subsection{Constant radii 
 circular strings in $AdS_5\times S^5$ }

It is straightforward to  combine the  solutions 
of sections 5.1 and 3.1 to write down the most general 
circular constant-radii solution in $AdS_5\times S^5$.
It will be parametrized by the 3+3 
 frequencies ($a=1,2; \ i=1,2,3$)
\be \la{fff}
\w_0=\k \ ,\ \ \ \ \ \ \w^2_a = k_a^2 + \k^2 \ , \  \ \ \ \ \ \ \ \  
w^2_i = m^2_i + \nu^2 \ ,  \ \ \ \ \ \k^2= \td \L \ , \ \ \ 
\nu^2 = -\L\ ,  \ee 
related to the 
energy $\E$ and 2+3 spins $\S_a $ and 
$\J_i$,    and by the topological numbers $k_a$ and $m_i$.
These will be  
related by \rf{kpp} and \rf{hjh} as well as by the conformal
gauge constraints \rf{cvi} and \rf{cvvi}. Explicitly, 
we get the following generalization of both \rf{onk}--\rf{uiy}
and \rf{cuat},\rf{nnb}
\bea\la{kqp}
 \ss {\J_i \ov  \sqrt{m^2_i + \nn^2} } =1 \  , \ \ \ \ \ \
\ \ \ \ {\E\over\kappa } - 
 \sum_{a=1}^2 {\S_a \over \sqrt{ k^2_a + \k^2}} = 1 \ , \\
\la{oiy}
 2\k \E -  2 \sum^2_{a=1} \sqrt{ k^2_a + \k^2 }\ 
 \S_a  - \k^2   =  2 \ss  \sqrt{m^2_i + \nn^2}\   \J_i  - \nn^2 \ ,
 \\
 \la{seo}
\sum_{a=1}^2   k_a \S_a  + \ss m_i \J_i =0\ . 
\eea
For given (integer or half-integer, in quantum theory) spins $S_a$ and 
$J_i$ the solution exists only for such integers 
$k_a$ and $m_i$ that satisfy \rf{seo}.
Assuming that all spins are of the same order and large 
$\S_a \sim \J_i \gg 1$ we find 
\bea \la{joj}
\k &=& \J  + { 1 \ov 2 \J^2 } ( \ss  m^2_i \J_i+2 \sa  k^2_a
\S_a)  + {\cal O}({1 \ov \J^2}) \ , \ \ \ \ \ \ \ \ \ 
   \J\equiv  \ss  \J_i \ , \\
\nu&=&\J  -   { 1 \ov 2 \J^2 }  \ss  m^2_i \J_i  
  + {\cal O}({1 \ov \J^2}) \ ,\eea
and thus 
\be \la{enna}
E = J  + S +  { \lambda \ov 2 J^2 } ( \ss  m^2_i J_i+ \sa  k^2_a S_a)  +
{\cal O}({\lambda^2 \ov J^3}) \ . \ee
This expression is a direct generalization of \rf{epon}
in the $\S_a=0$ case. 
The energy is minimal  if $m^2_i$ and $k_a^2$ 
have minimal possible values (0 or 1). 
We may also 
look at a different  limit when $ \J \gg \S \gg 1$
(corresponding to $k^2_1 \gg m^2_i$). 
In this case we get a ``BMN-type'' (single $J$ rotation type)
asymptotics with the leading term still given by 
\rf{joj}, i.e. $\Delta \E \sim { 1 \ov 2 \J^2 } \S$. 

 The conclusion is that  to have a 
regular (i.e. with analytic $\lambda $-dependence) 
large-spin expansion of the energy  one needs to have at
least one large  component of the spin in $S^5$ direction. 
This turns out to be the same also in the case of other
spinning string solutions with more complicated 
$\s$-dependence.

\bigskip

As an  explicit 
example, let us consider the simplest hybrid  solution
when only one of  each two types of spin  is  non-zero, 
i.e. $\J_1=\J, \ \S_1=\S, \  \ \S_2=\J_2=\J_3=0$.
The string then has $\rr_0^2 - \rr^2_1=1,\  \rr_2=0$ 
and $r_1=1,\ r_2=r_3=0$, i.e.  (cf. \rf{relx}) 
\be \la{jofj}
\YY_0 = \cosh \r_0 \ e^{i\k \tau} \ , \ \ \ \ 
\YY_1 = \sinh \r_0 \ e^{i\w \tau + i k \s}\ , \ \ \ \ \ \ \
\XX_1= e^{i w \tau + i m \s} \ , \ee
where $\rr_0= \cosh \r_0$ 
determines  the fixed radial coordinate in
$AdS_5$ at which the circular string is located while
 it is spread 
and rotating in $\phi_1$ (it is positioned  at  $\theta={\pi\ov 2}$
and $\phi_2=0$ in $S^3$ of  $AdS_5$). 
Also, the string is a rotating circle along $\vp_1$ in
$S^5$  located at $\vp_2=\vp_3=0, \ \g={\pi\ov 2}, \ \psi=0$. 
Its energy  for $\J \sim \S \gg 1$ is then \foot{Here $ \J = \sqrt{
m^2 + \nu^2}$, 
$2 \k \E - \k^2 = 2 \sqrt{ k^2 + \k^2} \S + \J^2 + m^2$ , 
$k \S + m \J=0$, \ $\E = \k + { \k \S \ov \sqrt{ k^2 + \k^2}} $.} 
\be \la{spl}
\E= \J + \S + { 1 \ov 2 \J^2} ( m^2 \J + k^2 \S) + ... 
=   \J + \S + { 1 \ov 2 \J} k^2 
{\S \ov \J} ( 1 + { \S\ov \J})  + ... \ , \ee 
where we used that $k \S + m \J=0$
and treat $ \S,\J$ and $k$  as independent data. 
Restoring $\l$ dependence  we thus  have\foot{The ``BMN-type''
    limit (cf. \ci{ft1,Russo}) 
    here corresponds to ${ \S\ov \J} \ll 1$.}  
\be \la{spli}
E= J + S + { \l k^2 \ov 2 J} {S \ov J} ( 1 + { S\ov J}) 
    + ...   \ . \ee 
It  should be possible to reproduce 
the same expression as a 1-loop anomalous dimension 
on the SYM side as was done for the folded $(S,J)$ solution 
in \ci{bfst}.

One can easily analyse the small  fluctuations near 
 this solution as was done in  section 3.4. 
One  finds 1 massless and 4 massive (mass $\nu$) fluctuations in
 $S^5$
 directions. In addition to 2 massive (mass $\k$) 
 decoupled $AdS_5$ fluctuations 
 there are also 3 coupled ones with a Lagrangian similar to
 \rf{uuu}:
 to obtain it  one is to do the following replacements in 
 \rf{uuu}:
 $f_2 \to  f_1,\ g_2\to g_1  ,  \ 
  f_1 \to  if_0,     \ w_1\to  \k,
  $ $\ w_2 \to \om_1=\sqrt{\k^2 + k^2}, \
   \ m_1=0, \ m_2=k, \ 
  a_2 \to i \rr_1, \  a_1 \to \rr_0,$ so that the equation \rf{bbb}
  for
  the characteristic frequencies $\omega$ 
   becomes
  \be 
  (\omega^2 -n^2)^2 + 4 \rr^2_1 ( \k \omega)^2  - 4 \rr^2_0 
  ( \sqrt{k^2 + \kappa^2}  \omega - k n)^2 =0 \ . \ee
The solutions of this equation 
  are  real. Indeed,   the analog of \rf{kuk} is
   found to be 
 \be \la{kuek}
 \Omega_{-} =  { 1\ov 2\k}
n \bigg[  2  (1+ \rr^2_1)  k 
\pm   \sqrt{n^2+  4   \rr^2_1 ( \rr_1^2+1)   k^2  } \bigg]
 + {\cal O}( { 1\ov \k^3}) 
 \ .  \ee 
 We conclude that 
   (in contrast to similar $(S_1,S_2)$ and
 $(J_1,J_2)$  circular  solutions) 
 this  hybrid $(S,J)$  solution is  always {\it stable}.

It should be possible  to  match \rf{spli} with 
 anomalous dimensions of particular 
  $\tr (D^S \Phi^J)+...$ operators 
on the SYM side by identifying 
the corresponding distribution of Bethe roots  in the 
Bethe ansatz equations of the 
associated  XXX$_{-1/2}$ Heisenberg spin chain \ci{bes2}, 
as was done for other folded and circular $(S,J)$  string 
solutions in \ci{bfst}.

\setcounter{equation}{0}
\section{Conclusions}

To summarize, we have found, in particular, 
 a solution of circular type with five spins 
$(S_1,S_2,J_1,J_2,J_3)$ whose leading
large-spin correction in the energy looks like 
 a one-loop 
term from the viewpoint
of SYM theory. Therefore,  it is plausible that
it can be matched onto the one-loop anomalous dimension
corresponding to certain 
 Bethe root distributions on the SYM side.
The string 
prediction for this  anomalous dimension is 
summarized  in
eq. (\ref{enna}) (with  (\ref{epo}) as a particular case).
Deriving it from the spin chain \ci{bes2} Hamiltonian 
 would clarify, in particular,  how the 
winding numbers of circular string states 
are encoded in the Bethe root distributions.

One interesting special 
case is that of  the solution with a single spin component $S$
in $AdS_5$ and a single R-charge $J$.
We have shown that this solution is stable for all values of spins and
winding numbers.
The corresponding energy formula   in (\ref{spli}) is 
very simple; it should be possible to  reproduce
 it on the SYM side as was done for other $(S,J)$ solutions in \ci{bfst}.
 

{}For general solutions of the Neumann-Rosochatius system,
the energy is a complicated implicit function of spins and 
topological numbers.  For example, in the two-spin
case of section  4.2  the general solution of the NR system can be
written in terms of elliptic functions but the 
energy  is a solution of  a parametric 
transcendental system of equations.
It would be very interesting to find a more direct
 map between the NR
system
and  Bethe equations for some properly 
chosen Bethe root distributions on the SYM side. 
It would also be important to find 
new pulsating solutions of the NR system mentioned in section 2.3
 that may have  simple SYM   counterparts.

\section*{Acknowledgments}
We are grateful to  N. Beisert, S. Frolov, H. Nicolai, M. Staudacher, 
 S. Theisen and K. Zarembo for discussions of some related issues.
We acknowledge partial support by
the European Commission RTN programme under 
contract HPNR-CT-2000-00131.
The work of  G.A. is supported in part
by RFBI grant N02-01-00695.
The  work of A.T.  is supported by the DOE grant
DE-FG02-91ER40690, and 
by the  INTAS  99-1590 grant and the RS Wolfson award. 
The work of J.R. is
supported in part by  MCYT FPA 2001-3598 and CIRIT GC 2001SGR-00065.



\begin{thebibliography}{20}

\bibitem{mz1}
J.~A.~Minahan and K.~Zarembo,
{\it ``The Bethe-ansatz for N = 4 super
Yang-Mills,''}
JHEP {\bf 0303}, 013 (2003)
{\tt hep-th/0212208}.


\bibitem{bes2}
N.~Beisert and M.~Staudacher,
{\it ``The N = 4 SYM integrable super spin chain,''}
{\tt hep-th/0307042}. 

\bibitem{bes3}
N.~Beisert, C.~Kristjansen and M.~Staudacher,
{\it ``The dilatation operator of N = 4 super Yang-Mills theory,''}
Nucl.\ Phys.\ B {\bf 664} (2003) 131,
{\tt hep-th/0303060};
N.~Beisert,
{\it ``The complete one-loop dilatation operator 
of N = 4 super Yang-Mills theory,''}
{\tt hep-th/0307015}.

\bibitem{ft2}
S.~Frolov and A.~A.~Tseytlin,
{\it ``Multi-spin string solutions in
\adss,''}
Nucl.\ Phys.\ B {\bf 668}, 77 (2003),
{\tt hep-th/0304255}.


\bibitem{ft3}
S.~Frolov and A.~A.~Tseytlin,
{\it ``Quantizing three-spin string
solution \mbox{in \adss'',}}
JHEP {\bf 0307}, 016 (2003),
{\tt hep-th/0306130}.


\bibitem{mz2}
N.~Beisert, J.~A.~Minahan,
M.~Staudacher and K.~Zarembo,
{\it ``Stringing spins and spinning
strings,''}
JHEP {\bf 0309}, 010 (2003),
{\tt hep-th/0306139}.

\bibitem{ft4}
S.~Frolov and A.~A.~Tseytlin,
{\it ``Rotating string solutions: AdS/CFT
duality in non-supersymmetric
sectors,''}
Phys.\ Lett.\ B {\bf 570} (2003) 96,
{\tt hep-th/0306143}.

\bibitem{afrt}
G.~Arutyunov, S.~Frolov, J.~Russo and A.~A.~Tseytlin,
{\it ``Spinning strings in \adss  and integrable systems,''}
{\tt hep-th/0307191}.


\bibitem{bfst}
N.~Beisert, S.~Frolov, M.~Staudacher and A.~A.~Tseytlin,
{\it ``Precision spectroscopy of AdS/CFT,''}
JHEP {\bf 0310}, 037 (2003),
{\tt hep-th/0308117}.

\bibitem{AS} G. Arutyunov and M. Staudacher, 
{\it ``Matching Higher Conserved Charges for 
Strings and Spins}, {\tt hep-th/0310182}.

\bibitem{mz3}
J.~Engquist, J.~A.~Minahan and K.~Zarembo,
{\it ``Yang-Mills duals for semiclassical strings on $AdS_5 \times S^5$,''}
{\tt hep-th/0310188}.


\bi {bmn}
D.~Berenstein, J.~M.~Maldacena and
H.~Nastase, {\it ``Strings in flat space and pp waves
from N =4 super Yang Mills,''}
JHEP {\bf 0204}, 013 (2002),
{\tt hep-th/0202021}.



\bibitem{gkp}
S.~S.~Gubser, I.~R.~Klebanov and
A.~M.~Polyakov,
{\it ``A semi-classical limit of the
gauge/string correspondence,''}
Nucl.\ Phys.\ B {\bf 636}, 99 (2002), 
{\tt hep-th/0204051}.

\bibitem{ft1}
S.~Frolov and A.~A.~Tseytlin,
{\it``Semiclassical quantization of
rotating superstring in \adss,''}
JHEP {\bf 0206}, 007 (2002),
{\tt hep-th/0204226}.


\bi{mina}
J.~A.~Minahan,
{\it ``Circular Semiclassical String Solutions on \adss,''}
Nucl.\ Phys.\ B {\bf 648}, 203 (2003),
{\tt hep-th/0209047}.

\bi{lup}
K.~Pohlmeyer,
{\it ``Integrable Hamiltonian Systems And Interactions Through Quadratic
Constraints,''}
Commun.\ Math.\ Phys.\  {\bf 46}, 207 (1976).

\bi{LP}
M.~Luscher and K.~Pohlmeyer,
{\it 
``Scattering Of Massless Lumps And Nonlocal Charges In The
Two-Dimensional Classical Nonlinear Sigma Model,''}
Nucl.\ Phys.\ B {\bf 137}, 46 (1978);
H.~Eichenherr and M.~Forger,
{\it ``On The Dual Symmetry Of The Nonlinear Sigma Models,''}
Nucl.\ Phys.\ B {\bf 155}, 381 (1979).

\bi{ogi}
A.~T.~Ogielski, M.~K.~Prasad, A.~Sinha and L.~L.~Wang,
{\it ``Backlund Transformations And Local Conservation Laws For
Principal Chiral Fields,''}
Phys.\ Lett.\ B {\bf 91}, 387 (1980).


\bibitem{BN}
B.M.~Barbashov and V.V.~Nesterenko,
{\it ``Relativistic String Model In A Space-Time Of A Constant
Curvature,}
Commun.\ Math.\ Phys.\  {\bf 78}, 499 (1981); 
``Introduction To The Relativistic String Theory,''
{\it  Singapore,  World Scientific} (1990) 249 p.



\bibitem{veg}
H.~J.~De Vega and N.~Sanchez,
{\it ``Exact integrability of strings in D-Dimensional De Sitter
space-time,''}
Phys.\ Rev.\  {\bf D47}, 3394 (1993).
%

\bi{vesa} H.~J.~de Vega, A.~L.~Larsen and N.~Sanchez,
{\it  ``Semiclassical quantization of circular strings in de Sitter and
anti-de Sitter space-times,''}
Phys.\ Rev.\ D {\bf 51}, 6917 (1995),
{\tt hep-th/9410219}.


\bi{ves}
H.~J.~de Vega and I.~L.~Egusquiza,
{\it ``Planetoid String Solutions in 3 + 1 
Axisymmetric Spacetimes,''}
Phys.\ Rev.\ D {\bf 54}, 7513 (1996), {\tt hep-th/9607056}.





\bi{Mandal}
G.~Mandal, N.~V.~Suryanarayana and S.~R.~Wadia,
{\it ``Aspects of semiclassical strings in $AdS_5$,''}
Phys.\ Lett.\ B {\bf 543}, 81 (2002),
{\tt hep-th/0206103}.

\bibitem{ben}
I.~Bena, J.~Polchinski and R.~Roiban,
{\it ``Hidden symmetries of the \adss\ superstring,''}
{\tt hep-th/0305116}.

\bibitem{Dolan}
L.~Dolan, C.~R.~Nappi and E.~Witten,
{\it ``A relation between approaches to integrability in superconformal
Yang-Mills theory,''}
JHEP {\bf 0310}, 017 (2003),
{\tt hep-th/0308089}.

\bibitem{Gorsky}
A.~Gorsky,
{\it ``Spin chains and gauge / string duality,''}
{\tt hep-th/0308182}.

\bibitem{Alday}
L.~F.~Alday,
{\it ``Non-local charges on \adss\ and pp-waves,''}
{\tt hep-th/0310146}.


\bibitem{Beisert:2003ys}
N.~Beisert,
{\it ``Higher loops, integrability and the near BMN
limit,''}
{\tt hep-th/0308074}; 
{\it ``The su(2$|$3) Dynamic Spin Chain,''}
{\tt hep-th/0310252}.


\bibitem{KP}
T.~Klose and J.~Plefka,
{\it ``On the Integrability of large N Plane-Wave Matrix Theory,''}
{\tt hep-th/0310232}.




\bibitem{Neumann} C. Neumann, {\it ``The problemate quodam mechanico, quod
ad primam integralium ultraellipticorum classem revocatur'',} J. Reine
Angew. Math. 56 (1859) 46; \\
O.~Babelon and M.~Talon,
{\it ``Separation Of Variables For The Classical And Quantum Neumann
Model,''}
Nucl.\ Phys.\ B {\bf 379}, 321 (1992),
{\tt hep-th/9201035}.


\bibitem{gag} 
T.S. Ratiu, {\it ``The Lie algebraic interpretation of the
complete integrability of the Rosochatius system'',} AIP Proceed. 88,
Amer. Inst. Physics, New York (1982), pp. 109;
L.~Gagnon, J.~P.~Harnad, J.~Hurtubise and P.~Winternitz,
{\it ``Abelian Integrals And The Reduction Method For An Integrable
Hamiltonian System,''}
J.\ Math.\ Phys.\  {\bf 26}, 1605 (1985);
C. Bartocci, G. Falqui and M. Pedroni, 
{\it ``A geometrical approach to the separability of the
Neumann-Rosochatius system'',} {\tt nlin.SI/0307021};
R.~Kubo, W.~Ogura, T.~Saito and Y.~Yasui,
{\it ``Geodesic flows for the Neumann-Rosochatius systems,''}
YITP-97-46;
J. Moser, ``Geometry of Quadrics'', Chern Symposium, Berkeley (1979), 
p.147.


\bibitem{Khan}
A.~Khan and A.~L.~Larsen,
{\it ``Spinning pulsating string solitons in \adss,''}
{\tt hep-th/0310019}.



\bibitem{dhn}
R.~F.~Dashen, B.~Hasslacher and A.~Neveu,
{\it ``The Particle Spectrum In Model Field Theories From
Semiclassical
Functional Integral Techniques,''}
Phys.\ Rev.\ D {\bf 11}, 3424 (1975).
\bibitem{FT} L.~D.~Faddeev and L.~A.~Takhtajan, Hamiltonian methods in the Theory
of Solitons, Springer-Verlag Berlin Heidelberg 1987.


\bibitem{M1}
J.~M.~Maillet,
{\it ``Hamiltonian Structures For Integrable 
Classical Theories From Graded Kac-Moody Algebras,''}
Phys.\ Lett.\ B {\bf 167} (1986) 401;
{\it ``Kac-Moody Algebra And Extended Yang-Baxter 
Relations In The O(N) Nonlinear Sigma Model,''}
Phys.\ Lett.\ B {\bf 162} (1985) 137.


\bibitem{AvM} M. Adler and P. van Moerbeke, 
Adv. Math. 38 (1980) 267.
 
\bibitem{AT}
J.~Avan and M.~Talon,
{\it ``Alternative lax structures for the
 classical and quantum Neumann model,''}
Phys.\ Lett.\ B {\bf 268} (1991) 209.



\bibitem{ZS}
V.~E.~Zakharov and Shabat, Functional Analysis and its Applications 
{\bf 12} (1978) 3;
V.~E.~Zakharov and A.~V.~Mikhailov,
{\it ``Relativistically Invariant Two-Dimensional Models In Field 
Theory Integrable By The Inverse Problem Technique,''}
Sov.\ Phys.\ JETP {\bf 47} (1978) 1017
[Zh.\ Eksp.\ Teor.\ Fiz.\  {\bf 74} (1978) 1953].



\bibitem{Russo}
J.~G.~Russo,
{\it ``Anomalous dimensions in gauge theories from rotating 
strings in \adss,''}
JHEP {\bf 0206}, 038 (2002),
{\tt hep-th/0205244}.



\end{thebibliography}
\end{document}